%% file: paper.tex
\documentclass[sigconf]{acmart}
\input{texheader}
\begin{document}

\input{defs}

\title{SIMD$^2$: A Generalized Matrix Instruction Set for Accelerating Tensor Computation beyond GEMM}
\author{Yunan Zhang}
\affiliation{%
  \institution{University of California, Riverside}
  \country{USA}
}
  \email{yzhan828@ucr.edu}
\author{Po-An Tsai}
\affiliation{%
  \institution{NVIDIA Research}
  \country{USA}
}
  \email{poant@nvidia.com}
\author{Hung-Wei Tseng}
\affiliation{%
  \institution{University of California, Riverside}
  \country{USA}
}
  \email{htseng@ucr.edu}
\date{}
\thispagestyle{plain}
\pagestyle{plain}
\input{abstract}

\maketitle
\input{introduction}
\input{applications}

\input{arch}

\input{model}
\input{methodology}
\input{result}
\input{related_work}
\input{conclude}

\input{acknowledgements}

\input{texfooter}

%% file: texheader.tex
\AtBeginDocument{%
  \providecommand\BibTeX{{%
    \normalfont B\kern-0.5em{\scshape i\kern-0.25em b}\kern-0.8em\TeX}}}

\usepackage{enumitem}
\usepackage{epsfig}
\usepackage{graphicx}
\usepackage{algorithm}
\usepackage{algorithmic}
\usepackage{amsmath}
\usepackage{listings}
\usepackage{xspace}

\usepackage{multirow}
\usepackage{booktabs}
\usepackage{tikz}

\newcommand{\hl}[1]{#1}

\def\BibTeX{{\rm B\kern-.05em{\sc i\kern-.025em b}\kern-.08em
    T\kern-.1667em\lower.7ex\hbox{E}\kern-.125emX}}

\input{macros}
\input{remark}
\remarktrue

\copyrightyear{2022}
\acmYear{2022}
\setcopyright{rightsretained}
\acmConference[To Appear: ISCA '22]{To Appear: the 49th Annual International Symposium on
Computer Architecture}{June 18--22, 2022}{New York, NY, USA}
\acmBooktitle{To appear in the 49th Annual International Symposium on Computer
Architecture (ISCA '22), June 18--22, 2022, New York, NY, USA}
\acmDOI{10.1145/3470496.3527411}
\setcopyright{none}

\lstdefinestyle{customc}{
  belowcaptionskip=1\baselineskip,
  breaklines=true,
  frame=L,
  xleftmargin=\parindent,
  language=C,
  showstringspaces=false,
  basicstyle=\footnotesize\ttfamily,
  keywordstyle=\bfseries\color{green!40!black},
  commentstyle=\itshape\color{purple!40!black},
  identifierstyle=\color[HTML]{0920C7},
  stringstyle=\color{orange},
}
\lstdefinestyle{customasm}{
  belowcaptionskip=1\baselineskip,
  frame=L,
  xleftmargin=\parindent,
  language=[x86masm]Assembler,
  basicstyle=\footnotesize\ttfamily,
  commentstyle=\itshape\color{purple!40!black},
}

%% file: macros.tex
\renewcommand{\em}{\it}

\newcommand{\ignore}[1]{}

\def\cfigure[#1,#2,#3]{
\begin{figure}
\begin{center}

\includegraphics[width=3.2in]{#1} 
 
\caption[]{#2
} \label{#3}
 
\end{center}
\vspace*{-0.3in}
\end{figure}}

\def\cfiguredouble[#1,#2,#3,#4]{
\begin{figure}
\begin{center}
\includegraphics[width=3in]{#1} 
\\(a)\\
\includegraphics[width=3in]{#2} 
\\(b)\\
\caption[]{#3} \label{#4}
\end{center}
\vspace*{-0.2in}
\end{figure}}

\def\cfiguretemp[#1,#2,#3]{
\begin{figure}
\vspace*{0mm}
\begin{center}

\includegraphics[width=3.5in]{#1} 
 
\vspace*{-3mm}\caption[]{#2
} \label{#3}
 
\vspace*{-5mm}
\end{center}
\vspace*{-2mm}
\end{figure}}

\def\wfigure[#1,#2,#3]{
\begin{figure*}[t]
\vspace*{0mm}
\begin{center}

\includegraphics[width=6.4in]{#1} 
 
\vspace*{-3mm}\caption[]{#2
} \label{#3}
 
\vspace*{-5mm}
\end{center}
\vspace*{-2mm}
\end{figure*}}

\def\threefigure[#1,#2,#3,#4,#5]{
\begin{figure*}
\vspace*{0mm}
\begin{center}

\begin{tabular}{ccc}
\includegraphics[width=2in]{#1} & \includegraphics[width=2in]{#2} &  \includegraphics[width=2in]{#3} \\
(a) & (b) & (c) \\
\end{tabular}

\vspace*{-3mm}\caption[]{#4
} \label{#5}

\vspace*{-5mm}
\end{center}
\vspace*{-2mm}
\end{figure*}}

\def\threefigureSC[#1,#2,#3,#4,#5]{
\begin{figure}
\vspace*{0mm}
\begin{center}

\begin{tabular}{ccc}
\hspace{-0.1in}\includegraphics[width=1.2in]{#1} &
\hspace{-0.15in}\includegraphics[width=1.2in]{#2} &
\hspace{-0.15in}\includegraphics[width=1.2in]{#3} \\
(a) & (b) & (c) \\
\end{tabular}
\caption[]{#4
} \label{#5}
\vspace*{-0.1in}
\end{center}
\vspace*{-0.2in}
\end{figure}}

\def\dcfigure[#1,#2,#3,#4,#5,#6]{
{
\begin{figure*}
\vspace*{0.2in}\
\begin{center}
\begin{minipage}[c]{3in}{
\includegraphics[width=3in]{#1} 
\vspace*{-3mm}\caption[]{#2} \label{#3} \
}\end{minipage}\hspace*{0.5in}\
\begin{minipage}[c]{3in}{
\includegraphics[width=3in]{#4} 
\vspace*{-3mm}\caption[]{#5}\label{#6} \
}\end{minipage}
\end{center}
\vspace*{-0.4in}\
\end{figure*}
}
}

\def\qcfigure[#1,#2,#3,#4,#5,#6]{
{
\begin{figure*}
\vspace*{0.2in}\
\begin{center}
\begin{minipage}[c]{3in}{
\includegraphics[width=3in]{#1} 
\vspace*{-3mm}
}
\end{minipage}\hspace*{0.5in}\
\begin{minipage}[c]{3in}{
\includegraphics[width=3in]{#2} 
\vspace*{-3mm}
}\end{minipage}

\begin{minipage}[c]{3in}{
\includegraphics[width=3in]{#3} 
\vspace*{-3mm}
}
\end{minipage}\hspace*{0.5in}\
\begin{minipage}[c]{3in}{
\includegraphics[width=3in]{#4} 
\vspace*{-3mm}
}\end{minipage}
\end{center}
\caption[]{#5}\label{#6}
\end{figure*}
}
}

\def\twfigureabc[#1,#2,#3,#4,#5]{
{
\tiny
\begin{figure}
\begin{center}
\begin{minipage}[c]{3.4in}{
\includegraphics[width=3.4in]{#1} 
\vspace*{-4mm}
}
\end{minipage}
\\(a)\\

\begin{minipage}[c]{3.4in}{
\includegraphics[width=3.4in]{#2} 
\vspace*{-4mm}
}\end{minipage}
\\(b)\\

\begin{minipage}[c]{3.4in}{
\includegraphics[width=3.4in]{#3} 
\vspace*{-4mm}
}
\end{minipage}
\\(c)\\
\end{center}
\vspace*{-6mm}
\caption[]{#4}
\label{#5}
\end{figure}
}
}

\def\twfigure[#1,#2,#3,#4,#5]{
{
\begin{figure}
\vspace*{0.2in}\
\begin{center}
\begin{minipage}[c]{6.5in}{
\includegraphics[width=6.5in]{#1} 
\vspace*{-3mm}
}
\end{minipage}

\begin{minipage}[c]{6.5in}{
\includegraphics[width=6.5in]{#2} 
\vspace*{-3mm}
}\end{minipage}

\begin{minipage}[c]{6.5in}{
\includegraphics[width=6.5in]{#3} 
\vspace*{-3mm}
}
\end{minipage}
\end{center}
\caption[]{#4}\label{#5}
\end{figure}
}
}

\def\dwfigure[#1,#2,#3,#4]{
{
\begin{figure*}
\vspace*{0.2in}\
\begin{center}
\begin{minipage}[c]{3.5in}{
\includegraphics[width=3.5in]{#1} 
\vspace*{-3mm}
}
\end{minipage}

\begin{minipage}[c]{3.5in}{
\includegraphics[width=3.5in]{#2} 
\vspace*{-3mm}
}\end{minipage}

\end{center}
\caption[]{#3}\label{#4}
\end{figure*}
}
}

\def\dssfigure[#1,#2,#3,#4,#5,#6]{
{
\begin{figure*}
\vspace*{0.2in}\
\begin{center}
\begin{minipage}[c]{4in}{
\includegraphics[width=4in]{#1}
\vspace*{-3mm}\caption[]{#2} \label{#3} \
}\end{minipage}\hspace*{0.5in}\
\begin{minipage}[c]{2in}{
\includegraphics[width=2in]{#4}
\vspace*{-3mm}\caption[]{#5}\label{#6} \
}\end{minipage}
\end{center}
\vspace*{-0.4in}\
\end{figure*}
}
}

\def\dsfigure[#1,#2,#3,#4,#5,#6]{
{
\begin{figure*}
\vspace*{0.2in}\
\begin{center}
\begin{minipage}[c]{3in}{
\includegraphics[width=3in]{#1}
\vspace*{-3mm}\caption[]{#2} \label{#3} \
}\end{minipage}\hspace*{0.5in}\
\begin{minipage}[c]{3in}{
\hspace*{0.5in}\
\includegraphics[height=3in]{#4}
\vspace*{-3mm}\caption[]{#5}\label{#6} \
}\end{minipage}
\end{center}
\vspace*{-0.4in}\
\end{figure*}
}
}

\def\dsyfigure[#1,#2,#3,#4,#5,#6]{
{
\begin{figure*}
\vspace*{0.2in}\
\begin{center}
\begin{minipage}[c]{2.5in}{
\includegraphics[height=2.5in]{#1}
\vspace*{-3mm}\caption[]{#2} \label{#3} \
}\end{minipage}\hspace*{0.5in}\
\begin{minipage}[c]{2.5in}{
\includegraphics[height=2.5in]{#4}
\vspace*{-3mm}\caption[]{#5}\label{#6} \
}\end{minipage}
\end{center}
\vspace*{-0.4in}\
\end{figure*}
}
}

\def\dyfigure[#1,#2,#3,#4,#5,#6]{
{
\begin{figure*}
\vspace*{0.2in}\
\begin{center}
\begin{minipage}[c]{3in}{
\includegraphics[height=3in]{#1} 
\vspace*{-3mm}\caption[]{#2} \label{#3} \
}\end{minipage}\hspace*{0.5in}\
\begin{minipage}[c]{3in}{
\includegraphics[height=3in]{#4} 
\vspace*{-3mm}\caption[]{#5}\label{#6} \
}\end{minipage}
\end{center}
\vspace*{-0.4in}\
\end{figure*}
}
}

\def\dyoldfigure[#1,#2,#3,#4,#5,#6]{
{
\begin{figure*}
\vspace*{0.2in}\
\begin{center}
\begin{minipage}[c]{3in}{
\epsfysize=2.0in\
\hspace{0.5in}\
\epsfbox{#1}
\vspace*{-3mm}\caption[]{#2} \label{#3} \
}\end{minipage}\hspace*{0.25in}\
\begin{minipage}[c]{3in}{
\epsfysize=2.0in\
\hspace{0.5in}\
\epsfbox{#4}
\vspace*{-3mm}\caption[]{#5}\label{#6} \
}\end{minipage}
\end{center}
\vspace*{-0.4in}\
\end{figure*}
}
}

\def\wpfigure[#1,#2,#3,#4]{
\begin{figure*}
\vspace*{4mm}
\begin{center}

\includegraphics[width=#4]{#1} 

\vspace*{-3mm}\caption[]{#2
} \label{#3}

\vspace*{-5mm}
\end{center}
\end{figure*}}

\def\wprfigure[#1,#2,#3,#4,#5]{
\begin{figure*}
\vspace*{4mm}
\begin{center}

\includegraphics[width=#4, angle=#5]{#1} 

\vspace*{-3mm}\caption[]{#2
} \label{#3}

\vspace*{-5mm}
\end{center}
\end{figure*}}

\def\DoubleFigureWSlide[#1,#2,#3,#4,#5,#6,#7,#8,#9]{
\begin{figure*}
\vspace*{#9}
\begin{center}
\begin{minipage}{#4}
\includegraphics[width=#4]{#1}
\vspace*{-3mm}\caption{#2
}\label{#3}
\end{minipage}
\hspace{2em}
\begin{minipage}{#8}
\includegraphics[width=#8]{#5}
\vspace*{-3mm}\caption{#6
}\label{#7}
\end{minipage}
\vspace*{-5mm}
\end{center}
\end{figure*}
}

\def\DoubleFigureW[#1,#2,#3,#4,#5,#6,#7,#8]{
\begin{figure*}
\vspace*{0in}
\begin{center}
\begin{minipage}{#4}
\includegraphics[width=#4]{#1}
\vspace*{-3mm}\caption{#2
}\label{#3}
\end{minipage}
\hspace{2em}
\begin{minipage}{#8}
\includegraphics[width=#8]{#5}
\vspace*{-3mm}\caption{#6
}\label{#7}
\end{minipage}
\vspace*{-5mm}
\end{center}
\end{figure*}
}

\def\DoubleFigureWHack[#1,#2,#3,#4,#5,#6,#7,#8]{
\begin{figure*}
\vspace*{0in}
\begin{center}
\begin{minipage}{3in}
\includegraphics[width=#4]{#1}
\vspace*{-3mm}\caption{#2
}\label{#3}
\end{minipage}
\hspace{2em}
\begin{minipage}{3in}
\includegraphics[width=#8]{#5}
\vspace*{-3mm}\caption{#6
}\label{#7}
\end{minipage}
\vspace*{-5mm}
\end{center}
\end{figure*}
}

\def\ddcfigure[#1,#2,#3,#4]{
\begin{figure*}[t]
\vspace*{0.2in}\
\begin{center}
\begin{minipage}[c]{3in}{
\includegraphics[width=3in]{#1} 
}\end{minipage}\hspace{0.5in}\
\begin{minipage}[c]{3in}{
\includegraphics[width=3in]{#2} 
}\end{minipage}\vspace*{-0.10in} \caption[]{#3}\label{#4}
\end{center}
\vspace*{-0.4in}\
\end{figure*}
}

\def\dddcfigure[#1,#2,#3,#4]{
\begin{figure*}
\vspace*{0.2in}\
\begin{center}
\begin{tabular}{cc}
\includegraphics[width=3in]{#1} &
\includegraphics[width=3in]{#2} \\
(a) & (b) \\
\end{tabular}\vspace*{-0.10in}\caption[]{#3}\label{#4}
\end{center}
\vspace*{-0.4in}\
\end{figure*}
}

\def\qqcfigure[#1,#2,#3,#4,#5,#6]{
\begin{figure*}[t]
\begin{center}
\begin{tabular}{cc}
\includegraphics[width=3.3in]{#1} &
\includegraphics[width=3.3in]{#2} \\
(a) & (b) \\
\includegraphics[width=3.3in]{#3} &
\includegraphics[width=3.3in]{#4} \\
(c) & (d) \\
\end{tabular}\vspace*{-0.10in}\caption[]{#5}\label{#6}
\end{center}
\vspace*{-0.4in}\
\end{figure*}
}
\def\qqcfiguresinglecol[#1,#2,#3,#4,#5,#6]{
\begin{figure}[t]
\begin{center}
\begin{tabular}{cc}
\hspace*{-0.2in}\includegraphics[width=1.8in]{#1} &
\hspace*{-0.2in}\includegraphics[width=1.8in]{#2} \\
\hspace*{-0.2in}(a) & 
\hspace*{-0.2in}(b) \\
\hspace*{-0.2in}\includegraphics[width=1.8in]{#3} &
\hspace*{-0.2in}\includegraphics[width=1.8in]{#4} \\
\hspace*{-0.2in}(c) & 
\hspace*{-0.2in}(d) \\
\end{tabular}\vspace*{-0.10in}\caption[]{#5}\label{#6}
\end{center}
\vspace*{-0.4in}\
\end{figure}
}

\def\ddcfiguresinglecol[#1,#2,#3,#4]{
\begin{figure}[t]
\begin{center}
\begin{tabular}{cc}
\hspace*{-0.1in}\includegraphics[width=1.8in]{#1} &
\hspace*{-0.1in}\includegraphics[width=1.8in]{#2} \\
\hspace*{-0.1in}(a) & 
\hspace*{-0.1in}(b) \\
\end{tabular}\vspace*{-0.10in}\caption[]{#3}\label{#4}
\end{center}
\vspace*{-0.4in}\
\end{figure}
}

\def\qqcfigureInAColumn[#1,#2,#3,#4,#5,#6]{
\begin{figure}[t]
\begin{center}
\includegraphics[width=3.3in]{#1} \\
(a)\\
\includegraphics[width=3.3in]{#2} \\
(b) \\
\includegraphics[width=3.3in]{#3} \\
(c) \\
\includegraphics[width=3.3in]{#4} \\
(d) \\
\vspace*{-0.10in}\caption[]{#5}\label{#6}
\end{center}
\vspace*{-0.5in}\
\end{figure}
}

\def\sixfigure[#1,#2,#3,#4,#5,#6,#7,#8]{
\begin{figure*}[t]
\vspace*{0.2in}\
\begin{center}
\begin{tabular}{cc}
\includegraphics[width=3in]{#1} &
\includegraphics[width=3in]{#2} \\
(a) & (b) \\
\includegraphics[width=3in]{#3} &
\includegraphics[width=3in]{#4} \\
(c) & (d) \\
\includegraphics[width=3in]{#5} &
\includegraphics[width=3in]{#6} \\
(e) & (f) \\
\end{tabular}\vspace*{-0.10in}\caption[]{#7}\label{#8}
\end{center}
\vspace*{-0.4in}\
\end{figure*}
}

\def\ddcfigureSlide[#1,#2,#3,#4,#5]{
\begin{figure*}
\vspace*{#5}\
\begin{center}
\begin{minipage}[c]{3in}{
\includegraphics[height=3in]{#1} 
}\end{minipage}\hspace{0.5in}\
\begin{minipage}[c]{3in}{
\includegraphics[height=3in]{#2} 
}\end{minipage}\vspace*{-0.10in} \caption[]{#3}\label{#4}
\end{center}
\vspace*{-0.4in}\
\end{figure*}
}

\def\dcfigureSingleCol[#1,#2,#3,#4]{
\begin{figure}
\begin{center}
\begin{tabular}{cc}
\hspace*{-0.2in}\includegraphics[width=1.5in]{#1} &
\hspace*{-0.2in}\includegraphics[width=1.5in]{#2} \\
\hspace*{-0.2in}(a) & 
\hspace*{-0.2in}(b) \\
\end{tabular}\vspace*{-0.10in}\caption[]{#3}\label{#4}
\end{center}
\vspace*{-0.4in}\
\end{figure}
}

\def\cxfigure[#1,#2,#3]{
\begin{figure}
\vspace*{4mm}
\begin{center}
 
\epsfxsize=2.5in\
\epsfbox{#1}\
 
\vspace*{-0.10in}\caption[]{#2
} \label{#3}
 
\vspace*{-5mm}
\end{center}
\vspace*{-2mm}
\end{figure}}

\newcommand{\x}{$\times$}

%% file: remark.tex
\newif\ifremark
\long\def\remark#1{
\ifremark%
        \begingroup%
        \dimen0=\columnwidth
        \advance\dimen0 by -1in%
        \setbox0=\hbox{\parbox[b]{\dimen0}{\protect\em #1}}
        \dimen1=\ht0\advance\dimen1 by 2pt%
        \dimen2=\dp0\advance\dimen2 by 2pt%
        \vskip 0.25pt%
        \hbox to \columnwidth{%
                \vrule height\dimen1 width 3pt depth\dimen2%
                \hss\copy0\hss%
                \vrule height\dimen1 width 3pt depth\dimen2%
        }%
        \endgroup%
\fi}

%% file: defs.tex
\newcommand{\mm}{mm$^2$}
\newcommand{\figtitle}[1]{\textbf{#1}}
\newcommand{\us}{$\mu$s}
\newcommand{\fixme}[1]{#1}
\newcommand{\adrian}[1]{{\color{green}\textbf{#1}}}
\newcommand{\laura}[1]{{\color{pink}\textbf{#1}}}
\newcommand{\joel}[1]{{\color{red}\textbf{#1}}}
\newcommand{\ameen}[1]{{\color{blue}\textbf{#1}}}
\newcommand{\arup}[1]{{\color{yellow}\textbf{#1}}}
\newcommand{\hungwei}[1]{{{#1}}}
\newcommand{\andrew}[1]{{{#1}}}
\newcommand{\hank}[1]{{\color{red}\textbf{#1}}}
\newcommand{\Bella}[1]{{\color{blue}\textbf{\textit{#1}}}}

\newcommand{\note}[2]{{\color{red}\fixme{$\ll$ #1 $\gg$ #2}}}
\newcommand{\myitem}[1]{\hspace*{-\parindent}\textbf{#1}\hspace*{\parindent}}
\newcommand{\speedup}[1]{S\x{}}
\newcommand{\CMO}[1]{CMO}
\newcommand{\SIMDD}[1]{SIMD$^2$\xspace}
\newcommand{\cmofull}[1]{compound matrix operations}


%% file: abstract.tex
\begin{abstract}
        Matrix-multiplication units (MXUs) are now prevalent in every computing platform.
        The key attribute that makes MXUs so successful is the semiring structure, which
        allows tiling for both parallelism and data reuse. 
        Nonetheless, matrix-multiplication is not the only algorithm with such attributes.
        We find that many algorithms share the same structure and differ in only the core operation; 
        for example, using add-minimum instead of multiply-add.
        Algorithms with a semiring-like structure therefore have potential to be accelerated by a general-purpose
        matrix operation architecture, instead of common MXUs.
        
        In this paper, we propose \SIMDD{}, a new programming paradigm to support generalized
        matrix operations with a semiring-like structure.
        \SIMDD{} instructions accelerate eight more types of matrix operations, in addition to matrix multiplications.
        Since \SIMDD{} instructions resemble a matrix-multiplication instruction,
        we are able to build \SIMDD{} architecture on top of any MXU architecture with minimal modifications.
        We developed a framework that emulates and validates \SIMDD{} using NVIDIA GPUs with Tensor Cores. Across
        8 applications, \SIMDD{} provides up to 38.59\x{} speedup and more
        than 6.94\x{} on average
        over optimized CUDA programs, with only 5\% of full-chip area overhead.
        
    \end{abstract}

%% file: introduction.tex
\section{Introduction}
\label{sec:introduction}
Matrices are essential data structures at the core of scientific computing,
data and graph analytics as well as artificial intelligence (AI) and machine learning
(ML) workloads. Due to the stagnating general-purpose processor performance
scaling and memory-wall problem~\cite{memory-wall},
a recent trend of efficient computing
on matrices focuses on building hardware accelerators. Famous examples include
NVIDIA's Tensor Cores~\cite{T4,nvidia-ampere}, Google's Tensor Processing Units
(TPUs)~\cite{TPU}, and the recent IBM Power 10 MMA unit~\cite{ibm-power10}.
The demand of matrix-multiplication accelerators is so
strong that the upcoming generations of Intel and ARM CPU processors will also
provide matrix extensions and integrate MXUs~\cite{IntelAMX,ArmSME}.

\ignore{
The emergence of artificial intelligence and machine learning (AI/ML)
accelerators has helped support the rapid growth of demands on AI/ML
applications, under the stagnating CPU performance scaling.  The core of AI/ML
accelerators are matrix processing units since in modern AI/ML models,
the majority of the computation can be represented as matrix multiplications
(e.g., CNNs~\cite{resnet}, Transformers~\cite{transformer}).  Famous examples include
NVIDIA's Tensor Cores~\cite{T4,nvidia-ampere}, Google's Tensor Processing Units
(TPUs)~\cite{TPU}, and the recent IBM Power 10 MMA unit~\cite{ibm-power10}.  The demand 
of matrix-multiplication accelerator is so
strong that the upcoming generations of Intel and ARM CPU processors will also
provide matrix extensions and integrate MXUs~\cite{IntelAMX,ArmSME}.
}

Compared with conventional SIMD processors (e.g., GPGPUs),
MXUs are more efficient in general matrix multiplication (GEMM) for two reasons.
First, GEMMs are easy to parallelize.
Each MXU can take tiles from input matrices and generate an output tile, and 
multiple MXUs can work together to form a larger GEMM accelerator, temporally or spatially.
Second, GEMMs have a higher compute intensity than vector operations (e.g., saxpy~\cite{blas}).
Such compute intensity alleviates the memory-wall issue in modern throughput-oriented SIMD processors
and allows architects to simply add more compute throughput to scale the performance of MXU with 
the same on-chip and off-chip bandwidth limitation.


Besides GEMM, a wide-spectrum of problems, including all-pair-shortest-path, minimum spanning
tree as well as graph problems, have matrix-based
algorithms/solutions share the same computation pattern.
They all follow a \textit{semiring-like structure} -- $A\: \oplus \: (B \: \otimes \:
C)$, where the
problem generates results (or intermediate results) by performing two-step operations
($\oplus$ and $\otimes$)
on three matrix inputs ($A$, $B$ and $C$). 
For example, \hungwei{dynamic programming methods for all-pair-shortest-path
problems using All Pairs Bellman-Ford or Floyd-Warshall algorithms} can be expressed in a semiring-like structure 
through having the $\otimes$ operator represent the
addition-based distance update operations~\cite{APSPSemiring,
generalizedGEMM}, and
the minimum operation replaces $\oplus$ operator.

\ignore{
Floyd-Warshall algorithm
is essentially one iteration of $min(C, A \otimes A)$, where $C$ is the current
shortest paths, $A$ is the original adjacency matrix and $\otimes$ represents the
addition-based distance update operations in Floyd-Warshall
algorithm~\cite{APSPSemiring, generalizedGEMM}. And
the minimum operation replaces $\oplus$ in the semiring-like structure. }

However, as modern MXUs are
highly specialized for just GEMM or convolutions, programmers must perform non-trivial algorithm
optimizations (e.g., mapping matrix multiplications to
convolutions~\cite{lu2021large,GPTPU}) to tailor these applications for
supported matrix operations. Besides, the resulting program
may still under-utilize MXUs as mapping the original set of matrix operations
to GEMMs that require changing the dataflow 
or data layout of the program before the actual computation can start. Finally, for problems including 
the \hungwei{dynamic programming algorithms}, existing MXUs cannot provide native support for the required $\oplus$
and $\otimes$ operations and have to fallback to SIMD processors (e.g., CUDA cores),
even though these algorithms share the semiring-like structure with GEMM.

To address these issues, this paper presents the \SIMDD{} architecture to enable more efficient 
matrix operations for a broader set of applications. \SIMDD{} provides a \emph{wider} set of
matrix-based operations that naturally fit the application demands and abstract these 
functions through an appropriate set of instructions. 
\SIMDD{}
reuses and extends the function of existing MXUs and data paths to minimize
the overhead in supporting additional matrix operations. 

The \SIMDD{} architecture brings the following benefits in accelerating matrix applications.
First, programmers or compilers can leverage the richer set of instructions that naturally maps to
common matrix operations without sophisticated code
transformations, which facilitates matrix-based programming. 
By performing more matrix operations with a minimum
number of instructions, the \SIMDD{} instructions
further reduce the control and data movement overhead over conventional SIMD instructions
by exposing a matrix-based abstraction. 
\ignore{
To fundamentally release the programmers' burden of transforming matrix algorithms, improve the
efficiency of MXUs and broaden the range of applications, the MXU
architecture needs to support a \emph{wider} set of \CMO{}s that naturally fits
the application demands and abstract these functions through an appropriate a hardware/software
interface. By supporting more \CMO{}s, the programmer or the programming
framework does not necessary need to significantly tailor to an existing
algorithm as a corresponding \CMO{} already presents. The resulting program
using these \CMO{}s also pays almost zero overhead in transforming input
matrices/tensors, uses of MXUs' hardware resources more efficiently, and most
importantly, achieves better performance. 

To explore the potential advantages
and trade-offs of extending MXUs to enable 
more efficient support for a wider spectrum of matrix/tensor applications,
this paper presents a set of \SIMDD{} instructions, each operating on
matrices and an architecture that implements the proposed \SIMDD{} instructions. 
}

As an initial step in this direction, our \SIMDD{} architecture introduces eight more types of 
instructions for matrix computation, including (1) min-plus, (2) max-plus, (3)
min-mul, (4) max-mul, (5) min-max, (6) max-min, (7) or-and, and (8)
plus-norm, in addition to existing mul-plus instructions. Similar to existing
hardware-accelerated GEMM operations, these instructions also take tiles of
matrices as inputs and update the resulting output tile.
Therefore, these instructions can easily share
the same infrastructure of an existing MXU, including
instruction front-end, memory, and register files. As these \SIMDD{} instructions all
follow the same data flow and computation pattern, they can also share the operand delivery structure
and simply require a modified data path to perform new operations. 
\ignore{
, (1) plus and minimum (\texttt{min-plus}) and
(2) L2 Distance (\texttt{L2D}), in addition to (3) matrix
multiplications and accumulations (\texttt{MMA}) that typical Tensor Cores
or matrix extensions in future Intel/Arm processors.}

As the necessary hardware support of \SIMDD{} resembles 
existing MXUs, a \SIMDD{} architecture 
can be implemented on top of any matrix-multiplication accelerators, either in
standalone application-specific integrated circuit (ASICs) or as processing
elements in CPUs or GPUs. This paper presents \SIMDD{} in the form of
extending GPU architectures as this allows us to leverage existing interface/front-end of 
GPU programming models and mature software stacks, and focus on the benefits of the \SIMDD{} model. 
On the other hand, since modern matrix-based applications still rely on non-matrix
operations to complete all computation tasks, this architecture also
offers better performance by avoiding data movements across system
interconnects and taking advantage of existing high-bandwidth memory hierarchy in GPUs. 

We evaluate the proposed \SIMDD{} architecture and hardware units through software emulation and hardware synthesis. 
We also made the emulation framework and hardware design publicly available
through a web-hosted repository\footnote{You may find the code repository at
\url{https://github.com/escalab/SIMD2}}. 
We demonstrate 8 applications that can naturally leverage 
these operations in their core algorithms. With the proposed \SIMDD{} MXUs,
these applications enjoy up to 38.59$\times$ speedup and more than 6.94\x{} speedup on average. 
Synthesis results show that over a conventional MXU that supports only multiply-and-accumulate,
\SIMDD{} MXU adds 69\% area overhead while supporting 8 different operations under the same clock period.
This area overhead is 5\% of the total chip area according to public die shot photos.

In presenting the \SIMDD{} architecture, this paper makes the following contributions. \\
(1) It identifies a set of matrix applications with semiring-like structure and reveals strong potential
in performance gain if \SIMDD{} support is available in hardware. \\
(2) It proposes \SIMDD{} architecture, programming model, instructions, and hardware units to accelerate semiring-like applications. \\
(3) It evaluates the performance benefit of the proposed \SIMDD{} architecture,
    and the cost of \SIMDD{} hardware units over a common MXU to demonstrate the opportunity of a \SIMDD{} programming paradigm.

%% file: applications.tex
\section{The Case for \SIMDD{}}
\label{sec:caseofsimd2}
\input{tableApplications}
The motivation of proposing \SIMDD{} for
matrix and tensor problems comes from two sources--
A family of matrix algorithms that share the same semiring pattern in computation,
and the emergence of GEMM accelerators designed around the semiring pattern.
Both motivate the need and the possibility of a single umbrella that covers a large set of matrix algorithms
to facilitate efficient use of hardware components. 

\ignore{
Matrix is one of the most intensively-used and fundamental mathematical objects
in solving real-world problems. 
This paper presents \SIMDD{} as a
general-purpose programming paradigm that allows programmers to easily
describe matrix-based programs and enables efficient hardware support. This
section will introduce the concept of \SIMDD{}. }

\subsection{The Commonality among Matrix Problems}
\label{sec:semiring}
\input{codeGemmVsApsp}
Matrices provide a natural mathematical expression for linear systems,
graphs, geometric transformations, biological datasets, and so on. 
In addition to data representations, many applications using matrices as inputs and outputs also share the 
same algebraic structure in their algorithms. This algebraic structure
contains two binary operators, $\oplus$ and $\otimes$. The $\oplus$ operator 
satisfies properties analogous to addition. The $\otimes$ operator is associative
and typically has a multiplicative identity element analogous 
to multiplication. 
In other words, a large set of matrix algorithms can be formalized as:\\
\vspace{-0.1in}
\[
D = C \oplus (A \otimes B)
\]
where $A$, $B$, $C$ are input matrices, $D$ is the output, as well as the
two customized operators, $\oplus$ and $\otimes$. 

The above algebraic structure is similar to a semiring, $(R, \oplus, \otimes)$, which contains a set $R$ 
equipped with two binary
operators, $\oplus$ and $\otimes$. The $\oplus$ operator in a semiring 
satisfies properties analogous to addition.
The $\otimes$ operator in a semiring has more restrictions as it must be associative, distributive as well as having a multiplicative identity
element. Since some algebraic structure of matrix problems is similar,
but not mathematically identical to semirings, we use the term \emph{semiring-like structure}
when referring to this identified algebraic structure.

\ignore{
In abstract algebra, a semiring, $(R, \oplus, \otimes)$, is a set $R$ equipped with two binary
operators, $\oplus$ and $\otimes$.
The $\oplus$ operator satisfies properties analogous to addition. The $\otimes$ 
operator is associative, distributive as well as having a multiplicative identity
element, in other words, analogous to multiplication. Semirings are similar
to rings in all aspects, except that semirings do not require negative elements. 
}

General matrix multiplication (i.e., GEMM) is one classic example that follows this structure.
To simplify the discussion, we use square matrices in the following
examples. 
Let $A$ be an
$N$ by $N$ matrix and $a(i,j)$ represent the ($i$, $j$)-entry of $A$. Then,
there also exists two other $n$ by $n$ matrice, $B$ and $C$,
where $b(i,j)$ and $c(i,j)$ represent the ($i$, $j$)-entries of $B$ and $C$,
respectively. 
General matrix multiplication consists of a set of computation for the ($i$, $j$)-entry of
the resulting matrix $D$, $d(i,j)$, where $d(i,j) = c(i, j) + \sum^{N}_{k=0} a(i,k) \times
b(k,j)$. Figure~\ref{code:gemmapsp}(a) illustrates the code example for matrix multiplication
with $N \times N$ matrices. 
The matrix multiplication therefore has a semiring-like structure where the $\oplus$ operates as
pair-wise addition for each pair of elements sharing the same coordinate $i, j$
on each side of the operator, matrix $C$ and the result of $A \otimes B$. 
The $\otimes$ operates as calculating the value of the ($i$, $j$)-entry in
the result matrix $D$ as $\sum^{n}_{k=1} a(i,k) \times b(k,j)$ for each $i, j, k$. With the
aforementioned common form, a matrix
multiplication problem is $D = C + A \times B$. 

Besides matrix multiplications, a
wide spectrum of algorithms, especially those for solving graph problems or
algorithms that leverage dynamic programming, can
also be formulated as a structure similar to matrix multiplications by customizing the $\oplus$ and $\otimes$ operators. 
For example, Figure~\ref{code:gemmapsp}(b) shows how the \andrew{inner loops of all-piars}
\hungwei{Bellman-Ford} algorithm \andrew{\cite{DBLP:books/x/E2019}}
for all-pairs-shortest-path (APSP) problem is similar to the semiring-like 
algebraic structure as GEMM (Figure~\ref{code:gemmapsp}(a)). Each iteration in
Line 4--5 of Figure~\ref{code:gemmapsp}(b) performs the computation
of $d(i,j) = min\{c(i, j),  min^{N}_{k=0}[c(i,k) + a(k,j)]\}$, where each $d(i,j)$, $c(i, j)$, or $a(i,j)$ represents
the ($i$, $j$)-entry of matrix $D$, $C$ or $A$, respectively. The $D$ matrix
is the result of temporal all-pairs distances after the iteration, $C$ is the
result from the last iteration,
and $A$ is the original adjacency matrix. 
Therefore, we can leverage the semiring-like structure to express the \andrew{all-pairs Bellman-Ford} algorithm
for the APSP problem by replacing the $\oplus$ operator with $min$ and the $\otimes$ operator
with $+$. The core loops become $D = C{} \text{ } min{} \text{ } (C + A)$. 


In addition to the APSP problem,
there are other algorithms amenable to such a semiring-like structure.
Table~\ref{table:semirings} illustrates a set of problems and their corresponding 
customizations of $\oplus$ and $\otimes$ operators in their algorithms.

Though a semiring-like structure can serve as a generic programming paradigm
for matrix problems, conventional approaches in solving matrix problems require the programmers
to transform matrix data into lower-ranked data representations (e.g.,
scalar numbers or vectors) and redesign algorithms on these data
representations to fulfill the programming paradigm that modern CPUs and
GPUs can support. Performance optimizations on programs solving these
problems is especially challenging as they are intensive in both computation
and data accesses on conventional processor architectures. 

\subsection{Hardware Support for Semiring-like Structure in GEMMs Accelerators}
\label{sec:accelerators}
\begin{figure}[t]
  \begin{centering}
      \includegraphics[width=0.9\linewidth]{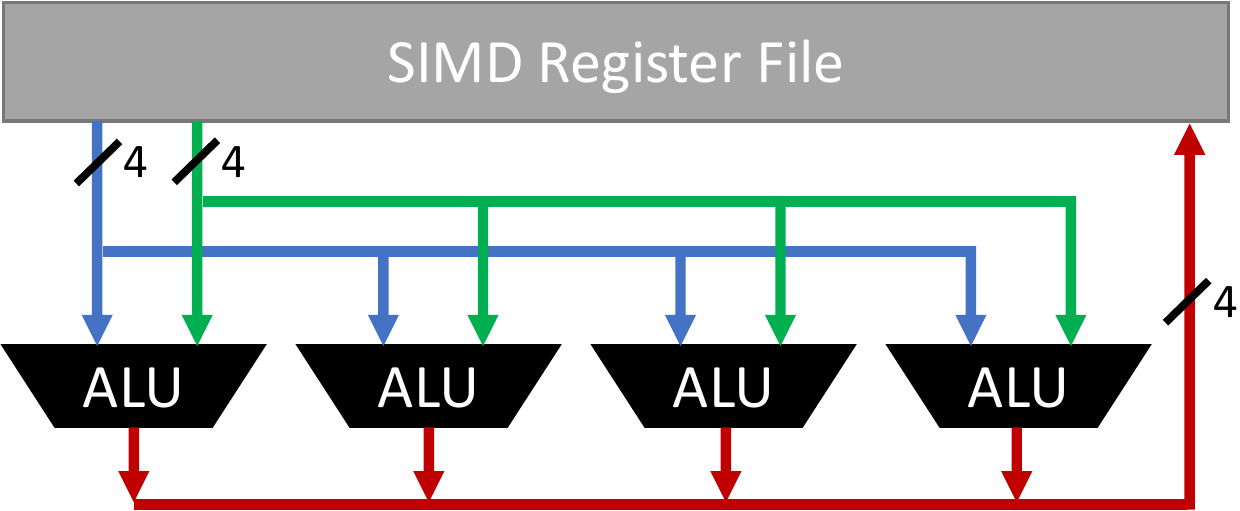}
      \vspace{-0.1in}
    \caption{\label{fig:simd_example}
        An example SIMD architecture.
    }
   \vspace{-0.2in}
  \end{centering}
\end{figure}

The semiring-like algebraic structure is the key enabler behind modern tensor accelerators, like MXUs for GEMMs,
which improves over conventional SIMD processors.
From a hardware design point of view, conventional SIMD architectures, shown in
Figure~\ref{fig:simd_example},
are bottlenecked by the vector register file bandwidth.
Such data transfer bottleneck (von Neumann bottleneck~\cite{von-neumann-bottleneck}) limits how many compute units (ALUs)
can be fed by the on-chip memory.
For example, a 4-wide register file can only supply to 4 ALUs at a time.
Even if the degree of parallelism grows as the problem size increases, the data transfer bottleneck remains.

\begin{figure}[t]
  \begin{centering}
      \includegraphics[width=0.9\linewidth]{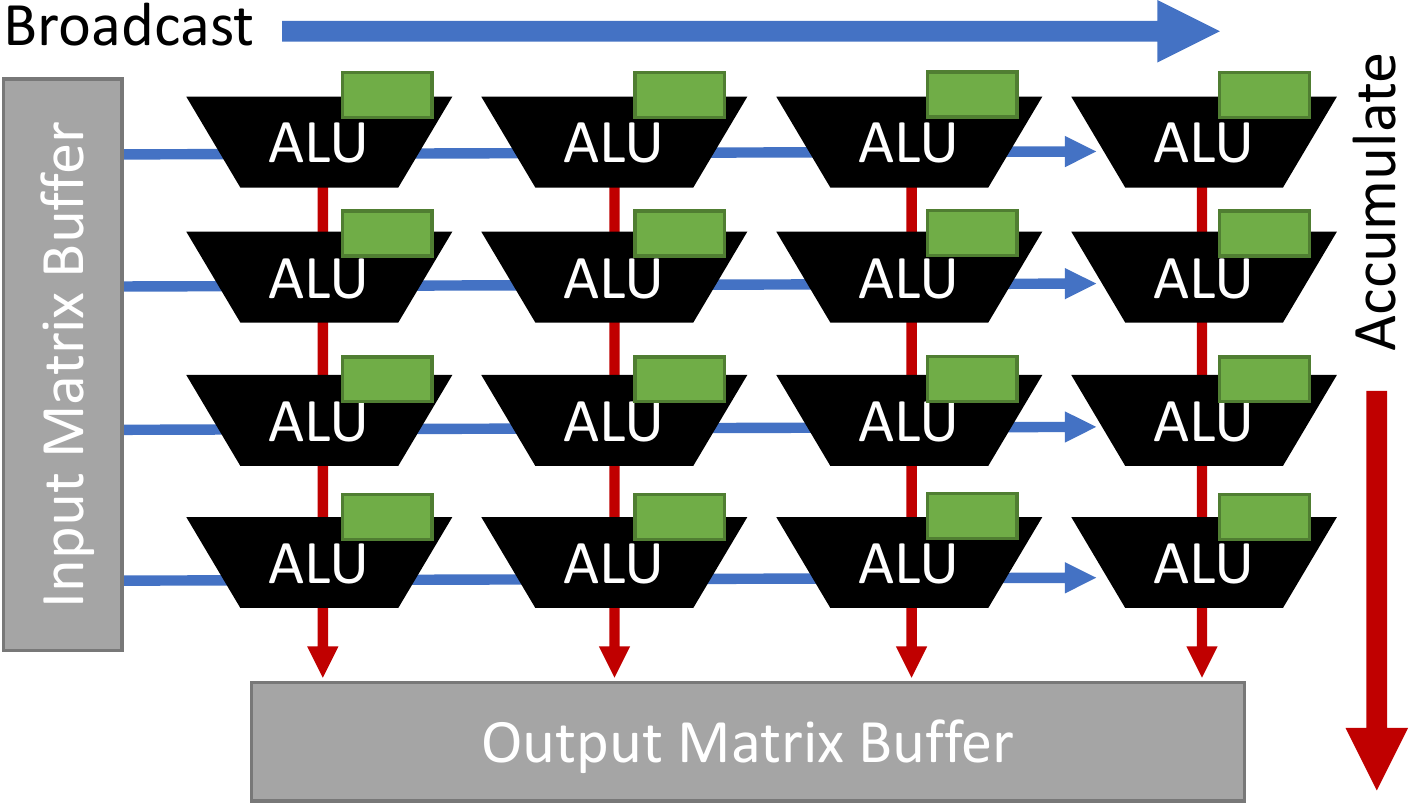}
      \vspace{-0.1in}
    \caption{\label{fig:mxu_example}
        An example MXU for GEMM.
    }
  \end{centering}
\vspace{-0.2in}
\end{figure}

MXUs, instead, leverage the semiring-like algebraic structure to break such bottleneck.
Figure~\ref{fig:mxu_example} shows an example implementation of MXU, modeled after the matrix unit in TPUs~\cite{TPU}.
In this MXU example, one input matrix is \emph{broadcast} to multiple ALUs because of the intrinsic data reuse opportunities in 
algorithms with a semiring-like structure.
The output matrix also leverages the structure (associative) and is accumulated across multiple ALUs before being stored into the output matrix buffer.
With the same 4-wide memory structure, we can now supply data to 16 ALUs.
More importantly, since the computation complexity is $O(N^3)$, and the data transfer is $O(N^2)$ in semiring-like algorithm,
the number of ALUs can scale much more than the on-chip memory bandwidth, alleviating the memory wall issue.

\begin{figure*}[t]
  \begin{tabular}{p{0.33\textwidth}p{0.33\textwidth}p{0.33\textwidth}}
  \multicolumn{3}{c}{\includegraphics[width=\textwidth]{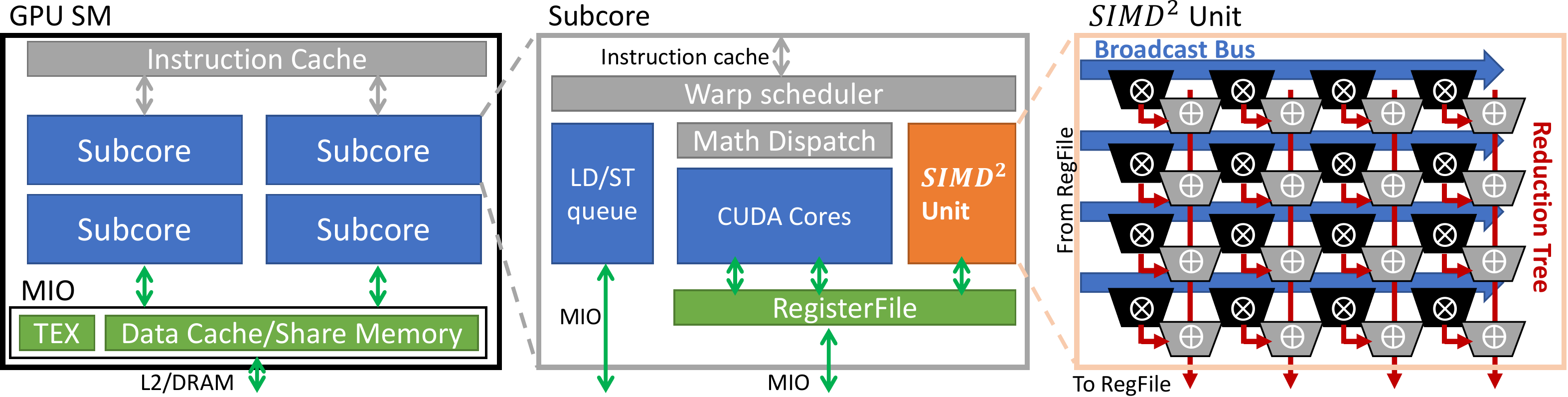}}\\
  \centering(a) & \centering (b) & \centering (c)\\
  \end{tabular}
  \vspace{-0.8cm}
      \caption[]{The high-level architecture of how \SIMDD{} units are integrated in GPU systems and the design of an \SIMDD{} unit.}
      \label{fig:simdd_arch}
  \vspace{-0.15in}
  \end{figure*}
  
As a result, modern MXUs are designed around the semiring-like structure, instead of optimizing the ALUs for multiply-add.
The programming model of these MXUs also leverages the nature of the algorithm to perform
work partitioning and tiling to execute a larger GEMM with multiple MXUs in a system~\cite{nvidia-ampere} or
across systems~\cite{tpuv3}.
For example, the \textit{wmma} API for NVIDIA Tensor Core works at the sub-tile granularity (e.g., 16x16),
and programmers can combine multiple \textit{wmma} calls to merge sub-tile into the full problem.

\textbf{Our insight}
is that supporting a wide range of semiring-like algorithms requires minimal changes on top of any systems with GEMM accelerators.
It is clear that the ALU in Figure~\ref{fig:mxu_example} is orthogonal to the hardware support (broadcast and accumulate)
for a semiring-like structure.
For example, if we enhance the ALU in Figure~\ref{fig:mxu_example} to support $add-minimum$, then
the same MXU architecture can now be used to accelerate solving APSP.
That is, the recent development of MXUs for GEMM has laid the ground of supporting semiring-like algorithms,
and with a better abstraction and hardware support, many more matrix algorithms can be accelerated.
This motivates us to propose and design \SIMDD{},
a new programming paradigm and architecture for semiring-like algorithms.

%% file: tableApplications.tex
\begin{table}
	\small
	\caption{Exemplary problems with their mappings to semiring-like structures and the corresponding
	definitions of operators to their solutions.}
	\centering
	\vspace*{-0.1in}
	\begin{tabular}{|l|c|c|l|}
	\hline
	Type of 	&	1st OP 	& 2nd OP 	&Representative \\
	matrix operations	&	$\oplus$& $\otimes$	&Algorithm(s) \\
	\hline
	Plus-Multiply	& $+$		&$\times$	& Matrix Multiplications, \\
			&		&		& Matrix Inverse	\\
	Min-Plus 	& $min$		& $+$		& All-pairs shortest paths\\
			&		&		& problem\\
	Max-Plus	& $max$		& $+$		& Maximum cost (critical
	path)\\
	Min-Multiply	& $min$		& $\times$	& Minimum reliability
	paths\\
	Max-Multiply	& $max$		& $\times$	& Maximum reliability
	paths\\
	Min-Max		& $min$		& $max$		& Minimum spanning tree\\
	Max-Min		& $max$		& $min$		& Maximum capacity paths\\
	Or-And		& $or$		& $and$		& Transitive and reflexive\\ 
			& 		&		& closure\\
	Add-Norm		& $+$		& $|a-b|^2$		& L2 Distance\\
	\hline
	\end{tabular}
	\label{table:semirings}
	\end{table}
	

%% file: codeGemmVsApsp.tex
    
    
\begin{figure*}[t]
\centering
\begin{minipage}[b]{0.45\linewidth}
    \centering
    \lstset{language=C,label=SliceExaple,basicstyle=\footnotesize}
        \begin{lstlisting}[frame=single, numbers=left, mathescape,label=scopingExample]
for (i = 0; i < N; i++)
  for (j = 0; j < N; j++)
    for(k = 0; k < N; k++){
      D[i][j] =  C[i][j]
        + A[i][k] * B[k][j];
    }
                       (a)
        \end{lstlisting}
        
\end{minipage}
\hspace{0.7cm}
\begin{minipage}[b]{0.45\linewidth}
    \centering
    \lstset{language=C,label=SliceExaple,basicstyle=\footnotesize}
        \begin{lstlisting}[frame=single, numbers=left,mathescape,label=scopesEnteredContents]
  for (src = 0; src < N; src++)
    for (dst = 0; dst < N; dst++)
      for(k = 0; k < N; k++){
        D[src][dst] =  min(C[src][dst],
                (C[src][k] + A[k][dst]));
      }
                       (b)
        \end{lstlisting}
        
\end{minipage}
\vspace{-0.1in}
    \caption{Code snippet of (a) GEMM and (b) APSP. See Section~\ref{sec:pm} for full APSP implementation.}
\label{code:gemmapsp}
\vspace{-0.1in}
\end{figure*}

\ignore{
\begin{minipage}[b]{0.45\linewidth}
    \centering
    \lstset{language=C,label=SliceExaple}
        \begin{lstlisting}[frame=single, numbers=left,mathescape,label=scopesEnteredContents]
while (!converge) {
  for (src = 0; src < N; src++){
    for (dst = 0; dst < N; dst++){
      for(k = 0; k < N; k++){
        D[src][dst] =  min(C[src][dst],
(C[src][k] + A[k][dst]));
      }
    }
  }
  converge = check_convergence(D, C);
}
        \end{lstlisting}
}

%% file: arch.tex
\section{\SIMDD{} Architecture}
\label{sec:simd2}
We propose the \SIMDD{} ISA to efficiently support matrix algorithms beyond GEMMs.
\SIMDD{} provides a programming paradigm and an instruction set to reflect the natural semiring-like
structure in solving these matrix problems.
The hardware units for \SIMDD{} instructions extend existing MXU to support the proposed programming
paradigm. This section will introduce both.

\subsection{The \SIMDD{} hardware architecture}
\label{sec:hardware}
\input{tableISA}
\input{tableSystemAPI}


Like GEMM accelerators, \SIMDD{} architecture can be implemented as a standalone processor that contains \SIMDD{} units only,
or functional units embedded with general-purpose scalar/vector processor cores to share the same instruction front-end. 
In this work, we chose the latter design and prototype \SIMDD{} architecture on a GPU as Figure~\ref{fig:simdd_arch} shows.
Specifically, we build on top of the NVIDIA SM architecture~\cite{choquette:hotchips18:volta},
which integrates Tensor Core as part of the subcore in a GPU SM.
The resulting high-level architecture resembles GPU SM
with Tensor Cores~\cite{raihan:ispass19:modeling-tensorcore} as the \SIMDD{} units implementing \SIMDD{}
instructions are part of a streaming multiprocessor, but the rest of the architectural components (front-end, memory-subsystem, etc.) are shared with conventional GPU cores.

\begin{figure}[t]
  \begin{centering}
      \includegraphics[width=\linewidth]{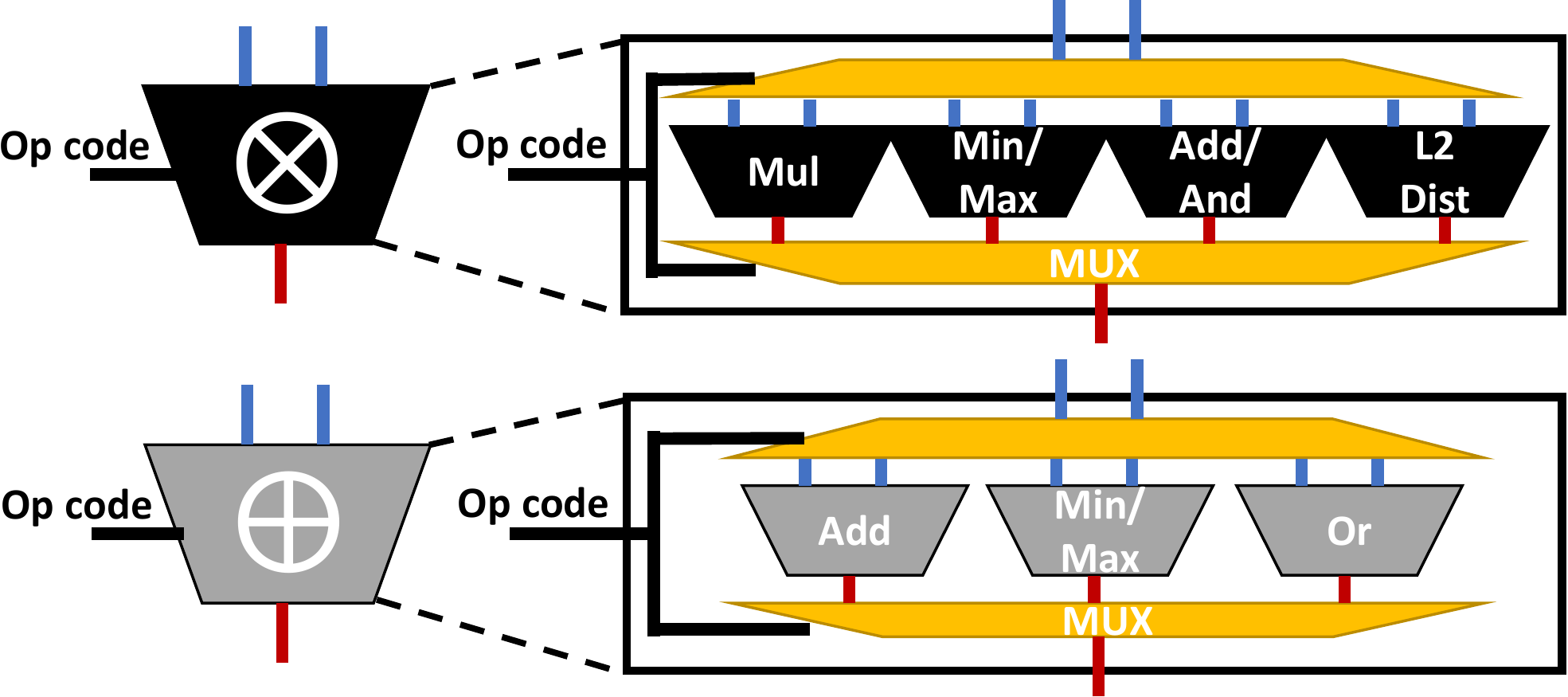}
      \vspace{-0.18in}
      \caption{\label{fig:simdd_alu}
        $\otimes$ ALU and $\oplus$ ALU in an \SIMDD{} unit.
    }
  \end{centering}
  \vspace{-0.22in}
\end{figure}

The \SIMDD{} unit in Figure~\ref{fig:simdd_arch}(c) extends conventional MXUs to use
different $\otimes$ and $\oplus$ operators.
Each \SIMDD{} unit can perform an \SIMDD{} arithmetic instruction using $\otimes$ operation on fixed-size 
matrix tiles (e.g., 4x4 in Figure~\ref{fig:simdd_arch}(c)) and produce an output matrix by reducing the result 
from $\otimes$ operation with the $\oplus$ operator.
Unlike tensor cores that only support multiply and accumulation,
the $\otimes$ ALU supports \textit{multiply}, \textit{min/max}, \textit{add/and}, and \textit{L2 dist},
and the $\oplus$ ALU supports \textit{add}, \textit{min/max}, \textit{or}, and \textit{subtract}.
Both ALUs are configured by decoding \SIMDD{} instructions, as shown in
Figure~\ref{fig:simdd_alu}. 

We chose to build \SIMDD{} architecture on top of GPUs for the following reasons.
First, since matrix operations just serve as the core computation in matrix applications,
applications typically rely on scalar or vector processors to preprocess or postprocess matrix data structures.
Collocating \SIMDD{} units with other processing elements enables efficient and fine-grained
data exchange and synchronization among heterogeneous computing units. 
Second, GPU's memory architecture design is more bandwidth-oriented and
serves better for the purpose since each \SIMDD{} unit would
consume/produce large amounts of data at once. Finally, there already exists
Tensor Cores in NVIDIA's GPU architecture that allow us to leverage
as a baseline design and an emulation framework. 

Alternatively, we have also explored implementing the \SIMDD{} unit by building a dedicated hardware unit for each semiring-like algorithm.
For example, in addition to the MXU for GEMM, we can add a hardware unit for \textit{min-add}, another unit for \textit{add-norm}, and so on.
Nonetheless, this design introduces 300\% area overhead (See Section~\ref{sec:area}) to the GEMM-only MXU,
which is $>4\times$ of the overhead introduced by the combined design in
Figure~\ref{fig:simdd_arch}. 

While we chose GPUs as the baseline system, building an \SIMDD{} architecture on other GEMM-based accelerators,
such as TPUs~\cite{TPU}, should be straightforward and low overhead.

\subsection{The \SIMDD{} ISA}
\label{sec:isa}
The \SIMDD{} instruction extension builds on top of the warp-level matrix-multiply-accumulate (\texttt{wmma}~\cite{wmma}) instructions for GPUs
and extends it to support new arithmetic instructions.
Table~\ref{table:ISA} lists these \SIMDD{} instructions.

The \texttt{load} instruction moves a chunk of data from the 1D shared memory address space as a fixed-size (16x16) matrix to the per-thread register file.
Like the \texttt{wmma} abstraction,
each thread in the warp stores part of the matrix in the register file and contributes to the whole warp-level operation.
The \texttt{store} instruction instead moves the matrix segments in the register file back to the 1D shared memory address space.

In our implementation, we assume input operands are always in 16-bit, half-precision floating-point format (fp16),
while the output data is always in 32-bit, single-precision floating point format (fp32).
While supporting other formats (e.g., \texttt{int8}) is possible, for many algorithms, 
we find fixed-precision format cannot converge to the same result as baseline fp32 implementations without \SIMDD{} instructions.

For the arithmetic operations,
we introduced eight more $\oplus$-$\otimes$ ops, in addition to the classic matrix-multiply-accumulate (\texttt{mma}).
These nine instructions map to the frequently used matrix problem patterns in Table~\ref{table:semirings}.
The \SIMDD{} arithmetic instruction shares the same register file as the vector processor,
and uses arguments that specify register locations of input and output matrices.
The latency of each \SIMDD{} instructions depends on the actual hardware implementation of the \SIMDD{} unit,
and in our implementation, we provision the \SIMDD{} unit to be the same throughput as the conventional MXUs
so that all \SIMDD{} arithmetic instructions have the same latency.

Similar to our changes for hardware architecture,
we expect adding the \SIMDD{} instructions to other ISAs that already support GEMMs,
such as Intel AMX~\cite{IntelAMX}, 
to be straightforward.
These matrix extensions already support matrices as input or output operands and provide data movement instructions for matrices (load/store matrix).
\SIMDD{} simply adds more arithmetic instruction on top of them.
We align our \SIMDD{} design point with modern GPU
architectures to facilitate our evaluation, but this is not fundamental.

%% file: tableISA.tex
\begin{table*}[h]
\small
\caption{A summary of the PTX instruction set architecture for \SIMDD{}}
\centering
\vspace*{-0.1in}
\begin{tabular}{|l|c|c|l|}
\hline
    \textbf{Data Movement Instructions}   & \textbf{Data Types}& \textbf{Matrix Shape} & \textbf{Source $\rightarrow$ Destination}\\
    \hline
    \SIMDD{}.load  & fp16 & 16x16 & Shared Memory $\rightarrow$ Register File\\
    \SIMDD{}.store & fp32 & 16x16 & Register File $\rightarrow$ Share Memory\\
\hline
    \textbf{Arithmetic Instructions} 	& 	\textbf{$\oplus$ OP} 	& \textbf{$\otimes$ OP} & \textbf{Algorithm}	 \\
\hline
    \SIMDD{}.mma	&  $+$		&$\times$	& GEMM\\
    \SIMDD{}.minplus 	& $min$		& $+$		& All-pairs shortest paths problem\\
    \SIMDD{}.maxplus	& $max$		& $+$		& Maximum cost (critical path)\\
    \SIMDD{}.minmul	& $min$		& $\times$	& Minimum reliability paths\\
    \SIMDD{}.maxmul	& $max$		& $\times$	& Maximum reliability paths\\
    \SIMDD{}.minmax	& $min$		& $max$		& Minimum spanning tree\\
    \SIMDD{}.maxmin	& $max$		& $min$		& Maximum capacity paths\\
    \SIMDD{}.orand	& $or$		& $and$		& Transitive and reflexive closure\\
    \SIMDD{}.addnorm	& $+$		& $|a-b|^2$		& L2 Distance\\

\hline
\end{tabular}
\label{table:ISA}
\end{table*}

%% file: tableSystemAPI.tex
\begin{table*}[t]
    \small
\centering
\caption{Example API of \SIMDD{} programming model}
\vspace{-0.1in}
\begin{tabular}{|p{0.4\textwidth}|p{0.55\textwidth}|}
\hline
\textbf{Sample Low-level Synopsis}                                                                                                                                                                          & \textbf{Description}                                                                                                                                 \\ \hline
simd2::matrix\textless{}matrix\_type, m, n, k, data\_type\textgreater{}                                                                                                                    & \begin{tabular}[c]{@{}l@{}}Declaration function,  declare the matrix will be applied in the m$\times$n$\times$k \\matrix-matrix operation.\end{tabular}   \\ \hline
simd2::fillmatrix(simd2::matrix, value)                                                                                                                                                    & Fill the target matrix with given value.                                                                                                             \\ \hline
simd2::loadmatrix(simd2::matrix, source, ld)                                                                                                                                               & \begin{tabular}[c]{@{}l@{}}Load value from source memory location to the target matrix, load with \\the step of leading dimension.\end{tabular}      \\ \hline
\begin{tabular}[c]{@{}l@{}}simd2::mmo(simd2::matrix,  simd2::matrix, \\                              simd2::matrix, simd2::matrix,                              simd2::opcode)\end{tabular} & \begin{tabular}[c]{@{}l@{}}Performs the matrix-matrix operation with given opcode.\end{tabular}                                                   \\ \hline
simd2::storematrix(target, simd2::matrix, ld)                                                                                                                                              & \begin{tabular}[c]{@{}l@{}}Store value to source memory location from the target matrix, store with \\the step of leading dimension.\end{tabular} \\ \hline
\multicolumn{2}{c}{\vspace{0.3cm}}\\
\end{tabular}
\label{table:sysapi}
\vspace{-0.15in}
\end{table*}

\ignore{
\hline
\textbf{Suggested High-level Synopsis}                                                                                                                                     & \textbf{Description}                                                                                                                                                                                                                                            \\ \hline
\begin{tabular}[c]{@{}l@{}}simdMM(int M, int N, int K, \\half * A, half * B, float * C,\\float * D, opcode op);\end{tabular}                                                               & \begin{tabular}[c]{@{}l@{}}This Function performs D = C + A * B, with size MxN, MxN, MxK, KxN respectively. \\ Choice of opcode can be chosen from min-plus, max-plus, plus-mul,max-mul, min-mul, \\ min-max, max-min, or-end, plus-norm.\end{tabular} \\ \hline
}

%% file: model.tex
\section{Programming Model}
\label{sec:pm}
The \SIMDD{} units in our proposed architecture can perform matrix
operations on a set of predefined matrix shapes and data types. Therefore,
the native programming interface reflects the abstraction by which these
\SIMDD{} units expose through the \SIMDD{} ISA. 
To further facilitate
programming at the application level, the framework can provide higher-level
library functions that decouple the programmability from
architecture-dependent parameters. 

\ignore{
On top of the instruction set and proposed architecture, this paper presents the programming model 
takes advantage of \SIMDD{} paradigm. To fully exploit the performance of \SIMDD{}, and increase 
the programbility, we suggest a format of high level API that can be implemented with previously 
mentioned instructions. Existing programming models exploit the integrated AI/ML accelerators 
providing easy-to-use software APIs, for instance, \textit{hgemm()} and \textit{gemmEX()} offered 
by Nvidia cuBlas can perform fast Gemm using Tensor Core; \textit{conv2D()} offered by tensorFlow 
can be accelerated with Google's TPU to directly accelerate 2-dimenional convolution. Other 
frameworks \cite{?} included cutlass, cuSparse also provided similar APIs that directly expose 
its core computation. Even though \SIMDD{} instructions are designed for low-level hardware, 
we suggest a format of high level API called \textit{simdMM()} which inherits the style of cuBLAS 
and cuASR. In this section, we will explain how to compose \textit{simdMM()} with proposed \SIMDD{} 
instructions, as well as how to use this high level API to implement listed applications mentioned in
Section~\ref{sec:caseofsimd2}.
}


Table~\ref{table:sysapi} summarizes the available functions from \SIMDD{}'s
low-level programming interface. 
Each of these functions maps directly to a set of 
instructions that Section~\ref{sec:isa} describes. The exemplary programming
interface resembles the C++ warp matrix operations that NVIDIA's Tensor Cores
use to smooth the learning curve, but not a restriction from the \SIMDD{}
architecture. 

Since the low-level interface reflects the architecture of \SIMDD{} units,
these functions must operate on a set of matrix shapes and data
types that the underlying \SIMDD{} hardware natively supports. 
The program needs to first declare the desired matrix shapes and
reserve the register resources for input matrices using the 
\texttt{simd2::matrix} function. Then, the program can load input matrices
into these reserved resources using the \texttt{simd2::loadmatrix} function or set values
using the \texttt{simd2::fillmatrix} function. The \texttt{simd2::mmo}
function receives arguments describing the desired \SIMDD{} operation to
perform on the input matrices and the location of the destination matrix.
After the code finishes necessary computation on these matrices, the \texttt{simd2::storematrix}
can reflect the updated values to a memory location. 
In case the source dataset does not fit the supported formats, the program
typically needs to explicitly partition datasets into tiles of matrices and aggregate
partial results appropriately.

To facilitate programming and alleviate the
burden of programmers, our framework provides
a set of high-level functions as an alternative programming interface. Each maps to
a specific type of \SIMDD{} arithmetic operations. These functions are
essentially composed using the aforementioned low-level functions. 
In contrast to the
low-level interface with limitations on inputs, these high-level functions
allow the programmer to simply specify the memory locations of datasets and
implicitly handle the tiling/partitioning of datasets and algorithms. 

\input{codeTiling}
Figure~\ref{code:tiling} provides an example code that implements a
high-level interface function that solves the \texttt{min-plus} matrix
problems. The compute kernel starts by identifying the logical \SIMDD{}
unit of the instance itself is occupying (Lines 6--7). The compute kernel then allocates
resources on the \SIMDD{} units (Lines 9--11). The code then loads the
current partial result of the target tile into one of the allocated matrix
storage (Line 13). The following for-loop (Lines 15--21) loads different pairs of tile
matrices from the raw input (Lines 17--18) and performs \texttt{min-plus} operations
(Line 20) on these tile matrices
together with tile loaded in Line 13. 

\input{codeAPIexample}
To use the compute kernel from Figure~\ref{code:tiling} or the low-level
\SIMDD{} interface, the programming model still requires a host program to
control the workflow, coordinate the computation on various types of
processors and move datasets among memory locations on heterogeneous
computing devices. 
Figure~\ref{code:apiexapmle} shows an example code that solves the all pair
shortest path problem using the \andrew{All-pairs Bellman Ford} algorithm. As \SIMDD{} units are auxiliary
computing resources to a GPU, the program code will need to explicitly allocate
GPU device memory (Line 4--10) and move data to the allocated space before invoking the high-level
\texttt{simd2\_minplus} function that Figure~\ref{code:tiling} implements (Line 14). 
\hungwei{The na\"ive SIMD$^2$ implementation of All-pairs Bellman Ford algorithm would require $V$ iterations of Line 14. 
The na\"ive implementation assumes the diameter of the graph is always the same as the number of vertices, the worst case scenario. 
However, the diameter of a real-world graph is way lower than that and a majority of iterations in Line 14 repeatedly generate identical results. 
Therefore, the implementation in Figure~\ref{code:apiexapmle} added a convergence check (i.e., the \texttt{check\_convergence} function call) in Line 15 to compare if any
element in the result matrix changes from the last iteration. If the result remains the same, the algorithm can terminate earlier. 
The \texttt{check\_convergence} (Line 15) is a pure GPU kernel. Because both
\SIMDD{} units and conventional GPU cores share the same device memory and
registers, the program does not need additional data movements between Line
14 and Line 15.
}

\hungwei{
In Figure~\ref{code:apiexapmle}, we use \andrew{All-pairs Bellman Ford} algorithm as the inputs of SIMD$^2$ computation in this algorithm are easier to understand.
In practice, the Leyzorek's Algorithm can solve APSP problem with fewer SIMD$^2$ operations~\cite{leyzorek1957investigation}.
Leyzorek's Algorithm still uses SIMD$^2$, but computes $C = C \oplus (C \otimes C)$ in Line 14 instead. 
In this way, Leyzorek's Algorithm only requires $lg |V|$ iterations to solve an APSP problem in the worst case scenario.
}

%% file: codeTiling.tex
\begin{figure}[t]
\lstset{language=c}          
\centering
\begin{minipage}[b]{0.85\linewidth}
\footnotesize
\lstset{language=C,label=SliceExaple,basicstyle=\footnotesize}
\begin{lstlisting}[frame=single, numbers=left, mathescape,label=scopingExample]
void simd2_minplus( half *A, half *B, 
                        float *C, float *D,
                        int m, int n, int k){
  // set tile ID
  int tile_id_y = get_tile_id_y();
  int tile_id_x = get_tile_id_x();
  // Declare simd2 matrices
  simd2::matrix<simd2::matrixa,16,16,16,half> mat_A; 
  simd2::matrix<simd2::matrixb,16,16,16,half> mat_B; 
  simd2::matrix<simd2::accum,16,16,16,float> mat_C;
  // load C to c_tile
  simd2::loadmatrix(mat_C, C, 16)
  // loop over K, each time do 16x16x16 mmo
  for(int tile_id_k=0;tile_id_k<k;tile_id_k+=16){
    // load A/B into a_tile/b_tile
    simd2::loadmatrix(mat_A, A, 16)
    simd2::loadmatrix(mat_B, B, 16)
    // performe mmo
    simd2::mmo(mat_C, mat_A, mat_B, mat_C, minplus);
  }
  // store back results
  simd2::storematrix(D,mat_C, 16);
} 
\end{lstlisting}
\end{minipage}
\vspace{-0.12in}
\caption{Tiled minplus MM on some architecture with \SIMDD{} supports}
\label{code:tiling} 
\vspace{-0.1in}
\end{figure}

%% file: codeAPIexample.tex
\begin{figure}[t]
\lstset{language=c}          
\centering
\begin{minipage}[b]{0.85\linewidth}
\lstset{language=C,label=SliceExaple,basicstyle=\footnotesize}
\scriptsize
\begin{lstlisting}[frame=single, numbers=left, mathescape,label=scopingExample]
float * adj_mat_d; 
float * dist_d_delta; 
float * dist_d;
cudaMalloc(..., adj_mat_d, ...);
cudaMalloc(..., dist_d, ...);
cudaMalloc(..., dist_d_delta, ...);

cudaMemcpy(adj_mat_d, ..., H2D);
cudaMemcpy(dist_d_delta, ..., H2D);
cudaMemcpy(dist_d, ..., H2D);

bool converge = true;  
while(converge){
    simd2_minplus(adj_mat_d, dist_d, dist_d, dist_d_delta, v, v, v);
    converge = check_convergence(dist_d, dist_d_delta, ...); 
}
cudaMemcpy(... , dist_d, ... , D2H);
\end{lstlisting}

\end{minipage}
\vspace{-0.15in}
\caption{CUDA kernel implenmentation of APSP using \SIMDD{} API}
\label{code:apiexapmle}
\vspace{-0.25in}
\end{figure}

\ignore{
// original graph adj matrix
float * adj_mat_d; 
// execution result of previous run.
float * out_d_delta; 
// new dist matrix after latest execution
float * out_d; 

// Memory allocation
cudaMalloc((float**)&adj_mat_d,
                v*v*sizeof(float));
cudaMalloc((float**)&out_d, 
                v*v*sizeof(float));
cudaMalloc((float**)&out_d_delta, 
                v*v*sizeof(float));

// Data movements
cudaMemcpy(adj_mat_d, adj_mat, 
                v*v*sizeof(float), cudaMemcpyHostToDevice);
cudaMemcpy(out_d_delta, adj_mat, 
                v*v*sizeof(float), cudaMemcpyHostToDevice);
cudaMemcpy(out_d, adj_mat, 
                v*v*sizeof(float), cudaMemcpyHostToDevice);

// convergence indicator
bool run = true;    

// stop at convergence
while(run){        
    // 1 iteration of minplus mm
    int retval = simdMM(
                    // shape of imput matrices    
                    v, v, v,
                    // maxtrix A        
                    adj_mat_d, v,
                    // maxtrix B
                    out_d, v, 
                    // maxtrix C         
                    out_d, v,
                    // maxtrix D        
                    out_d_delta,
                    // operator       
                    minplus);         
    // check convergence
    run = check_convergence(out_d, out_d_delta,v,v); 
}
//store computation result back to host
cudaMemcpy(dist, out_d, 
                v*v*sizeof(float), cudaMemcpyDeviceToHost);
}

%% file: methodology.tex
\section{Experimental Methodology}
\label{sec:methodology}
As \SIMDD{} promotes matrix-based algorithms, the \SIMDD{}-ized implementations of our 
benchmark applications may use different algorithms compared to their state-of-the-art 
implementations, typically using vectorized or scalar-based algorithms, on alternative 
platforms. Therefore, we designed a framework that allows us to validate the correctness 
of \SIMDD{}-ized programs and emulate the performance of \SIMDD{}-ized programs with or 
without \SIMDD{} hardware acceleration presented.
This section will describe these aspects in detail. 

\subsection{Emulation framework}
\label{sec:emu_validate}
To evaluate \SIMDD{}, we developed a framework that evaluates the correctness and performance for each
program under test on top of a testbed using a state-of-the-art GPU architecture.

\subsubsection{Hardware configuration}
\label{sec:exp_hardware}
Our validation and emulation framework uses a machine with NVIDIA's RTX 3080 
GPU based on the Ampere architecture with 10 GB device
memory. This machine has an 8-core, 16 threads AMD RyZen 3700X processor with
peak clock rate at 4.4~GHz and 16 GB physical main memory installed. The
machine hosts an Ubuntu 20.04~(Linux kernel version 5.13) with NVIDIA's CUDA
11.1 using driver version 470.103.01. 

\begin{figure}[t]
    \begin{center}
    \includegraphics[width=0.5\textwidth]{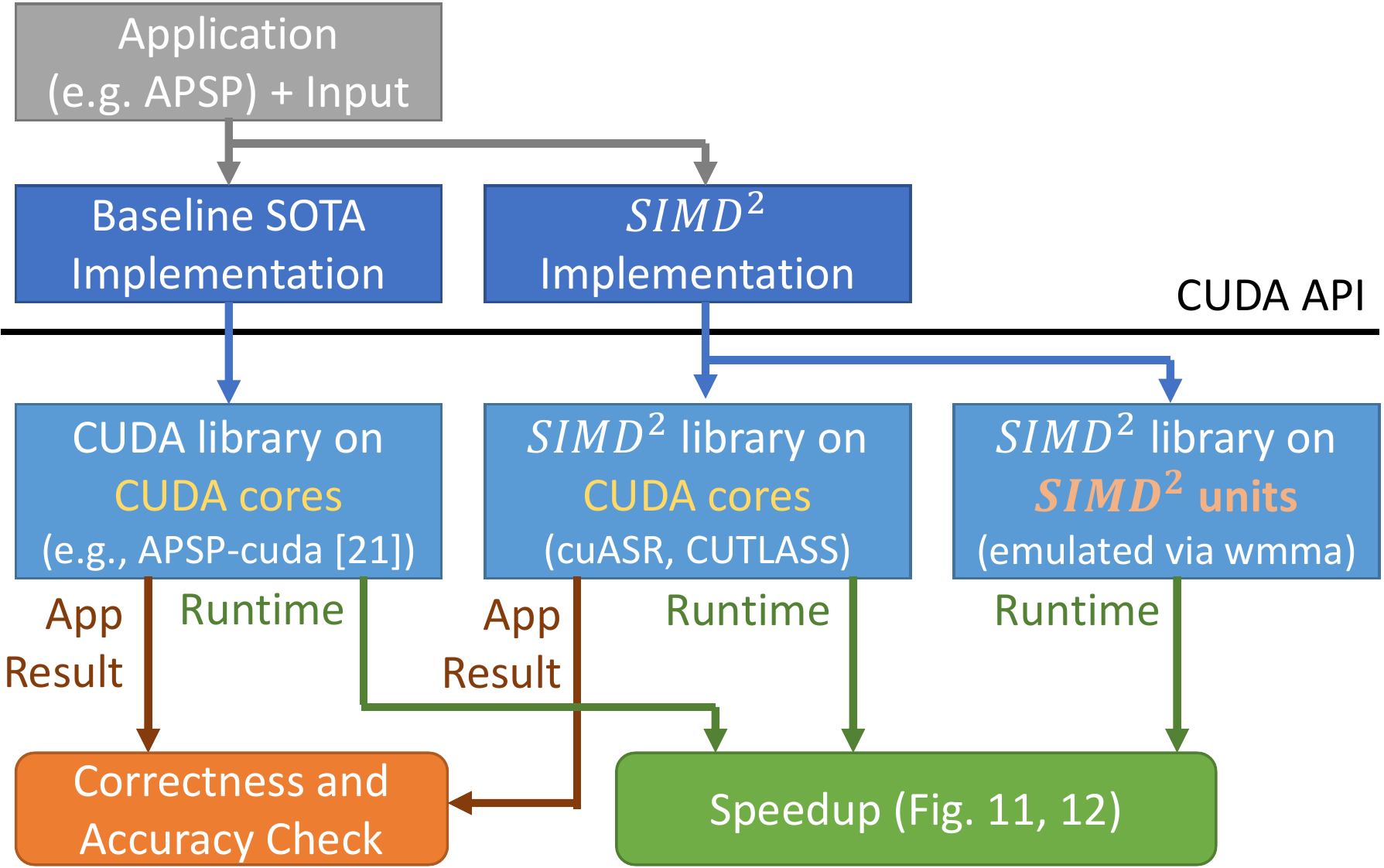}
    \vspace{-0.2in}
        \caption[]{The workflow of the emulation framework for \SIMDD{} evaluation.}
    \label{fig:eval}
    \vspace{-0.1in}
    \end{center}
\end{figure}

\subsubsection{Evaluation Process}
\label{sec:process}

\hl{Figure~\ref{fig:eval} illustrates the workflow of our process in evaluating applications. 
For baseline applications, we executed each application directly on the hardware platform 
without any modification to their source code, datasets and invoked library functions. For \SIMDD{}-ized 
applications, their implementations leverage semiring-like algorithms using the programming model 
and \SIMDD{} API functions described in Section~\ref{sec:pm}. The evaluation framework takes three types of inputs: (1) 
the compiled program and command line arguments, (2) the dataset used in the baseline application, and (3) the output of 
the baseline application with the input dataset and command line arguments. 
Once the emulation framework receives these three sets of inputs, the emulation framework can dynamically change the 
linked library that implements \SIMDD{} API functions to perform (1) correctness validation by using a backend that 
leverages conventional vector processors and compare the output with the output that the baseline version produced, or (2)
performance emulation by using a backend that generates instructions to Tensor Cores residing on the hardware platform. 
The following paragraphs will describe the correctness validation and performance emulation process in detail.}

\paragraph{Correctness validation}
\label{sec:correctness_validate}
In this work, we need to validate correctness in addition to performance emulation for the following reasons. 
First, as we need to alter the compute kernels to efficiently use \SIMDD{} units and \hl{in many cases,
using a different algorithm (e.g., Semiring-based vs. Kruskal's Algorithm in Minimum Spanning Tree problems)}, 
we need to verify if the change of implementation still delivers the same outcome as the baseline 
implementation. Second, \hl{as existing hardware accelerators only support MMA operations that cannot generate the correct 
output for other \SIMDD{} operations this paper proposes to extend, we need to verify if implemented semiring-based 
algorithms can generate the desired output after mapping the computation into the proposed \SIMDD{} units.}
Finally, \hl{this process can help collect the statistics regarding the total amount of various matrix 
operations and provide the input for performance emulation.}

During the validation process, we linked the backend of the \SIMDD{}
programming interface to a library that we extended from cuASR~\cite{cuASR}.
This library implements exactly the same functionality as the proposed
low-level \SIMDD{} functions, except that the library can simply leverage
CUDA cores through NVIDIA's high-performance CUTLASS library, but not use Tensor Cores. 
\hl{When implementing low-level \SIMDD{} functions for validation purposes, we carefully partitioned 
the inputs and outputs to fit the exact shape of matrix inputs and outputs of proposed \SIMDD{} units 
(i.e., the input/output sizes of each Tensor Core in our testbed) when invoking corresponding \SIMDD{} function calls. 
We also used reduced/mixed precision inputs/outputs to match the data types that our \SIMDD{} units support. 
Therefore, the validation process can help us access the accuracy of \SIMDD{} units.} 
For each program under test, we can optionally count the 
number of iterations, threads, and low-level \SIMDD{} function calls that are necessary 
to finish running the program and compare each program's output with its
state-of-the-art implementation on the alternative architecture. 

\paragraph{Performance emulation}
\label{sec:perf_eval}
The design of \SIMDD{} allows this work to leverage existing Tensor 
Cores that are available on the GPU of our emulation hardware for exact 
performance evaluation for the two main reasons. First, adding \SIMDD{} 
instructions do not increase the timing of an existing MMA unit 
(e.g., a Tensor Core) as Section~\ref{sec:area} 
reports. Second, the low-level instructions, register files, memory
hierarchy as well as the interaction with the host machine can be made almost
identical to those of Tensor Cores, except for the exact output after each
computation. 

\hl{When performing performance emulation, the framework links the backend of the 
low-level \SIMDD{} API library that implements through using equivalent 
Tensor Cores' WMMA low-level interface. As this paper simply proposes to extend the ALU functions of Tensor Cores, the 
memory operations remain the same in \SIMDD{} units compared with Tensor Cores. Therefore, each \texttt{simd2::loadmatrix} and \texttt{simd2::storematrix}
invocation are identical in its counterpart in CUDA's WMMA API.}
However, since \hl{the state-of-the-art Tensor Cores can only
perform MMA operations, the performance emulation backend library maps each invocation of \texttt{simd2::mmo} to a CUDA's 
\texttt{WMMA::mma} function call on the same size of inputs. This is also the main reason why the performance emulation backend
cannot produce correct/meaningful computation outcomes. The performance emulation process can optionally receive statistics 
from the corresponding validation process to compare if the performance emulation backend generates the exact amount of simd2 and 
WMMA operations as desired.}
This performance emulation methodology is
similar with prior work in extending Tensor Cores~\cite{EGEMM-TC} to support different precisions. 

\ignore{
In order to obtain the maximum performance of the tensor core, we chose to emulate the matrix 
operations supported by SIMD2 using \textit{gemmEX()} provided by cuBLAS. Since using a simple 
matrix multiplication kernel is not enough to obtain correct results and we cannot control 
the convergence of the computation, we use the number of iterations run by the cuASR kernel 
to control the number of gemmex runs, which emulates the performance of selected application 
after using \SIMDD{} API.
}
\ignore{
In order to demonstrate the correctness of \SIMDD{}, especially after applying the common optimization method of 
matrix multiplication. We applied each set of operators on the host kernel and the GPU simulated \SIMDD{} kernel 
to test the correctness on different shapes of matrices. We believe that the correctness benchmarks of the \SIMDD{} 
API are sufficient to demonstrate that 1. the \SIMDD{} API can inherit conventional matrix multiplication optimizations, 
whether for GPUs, CPUs, or special hardware. 2. the state-of-the-art GEMM kernel can be used to emulate the expected 
performance of the \SIMDD{} API. As a state-of-the-art software implementation of \SIMDD{}, the same correctness 
benchmark included cuASR kernels to ensure that the semiring matrix multiplication or semiring-liked matrix 
multiplication provided by cuASR is consistent with our \SIMDD{} API. Because the cuASR kernels were used 
extensively as designing performance benchmarking and implenmenting application of this work, we ensured the 
correctness of the cuASR kernels in the initial tests. Since the algorithms used in a large number of applications 
are in nature, only terminate at convergence, cuASR will be used as a correctness criterion for emulation and as a 
reference for the performance kernel to terminate its computation.
}
\ignore{
In this section, we introduce the experimental methodology of evaluating \SIMDD{} 
instructions in terms of precision and performance using existing hardware(software?) 
emulations. As discussed in the previous section, Tensor cores in modern Nvidia GPUs 
shares the same hardware abstraction as potential microarchiture that supports \SIMDD{} 
instructions, thus, we conducted emulation-based experiments on top of the foundation of Nvidia 
Tensor core architecture. This section will be divided into 5 subsections, (1) Experimental 
kernels, which introduces all kernel functions we applied to fully examine the performance and 
programmability of APIs implemented based on \SIMDD{} instructions. (2) Correctness and precision, 
which explains the software verification of each kernel mentioned. (3) Microbenchmark, which gives 
a overview of the mircobenchmarking of APIs implemented with \SIMDD{} instructions (4) Application 
benchmark, presents the detail of baseline and \SIMDD{} version of each applications.
\input{implementation}
}

\subsection{Applications}
\label{sec:benchmark}
\input{tableBaseline}
To demonstrate the performance of \SIMDD{}, we ran two types of workloads on the 
aforementioned evaluation framework. The first type is
a set of microbenchmark workloads that only iteratively invoke \SIMDD{}
functions and accept synthetic datasets to
help us to understand the pure
performance gain of \SIMDD{} instructions over alternative implementations.

The other is a set of full-fledged benchmark applications where each program
contains not only \SIMDD{} functional, but also interacts with other types
of processors to complete the tasks. These benchmark applications can accept 
real-world datasets and generate meaningful outputs accordingly for us to
assess the quality of results if appropriate. 

For each workload, we evaluate three implementations. 
\ignore{
\paragraph{State-of-the-art CPU implementation} This version of workload
implements the desired computation with the most efficient CPU code. Due to
the limited parallelism on modern CPUs, the performance is typically not highly-competitive
even with state-of-the-art algorithms. However, this implementation provides
a reference for a correctness validation process in our framework. }

\paragraph{State-of-the-art GPU baseline} This version of code serves as the
baseline of our workloads. We tried our best to collect implementations from publicly
available open-source code hosting websites and select the best-performing
implementation on our testbed as the state-of-the-art baseline version for
each workload. These implementations simply leverage CUDA cores, but not
Tensor Cores to accomplish their tasks. In fact, without a work like
\SIMDD{}, none of the selected benchmark can leverage Tensor Cores due to
the limited MMA functions available on such hardware units. 

\paragraph{\SIMDD{} in CUDA cores} This version of code serves as
another baseline of our workloads. This set of programs implement \SIMDD{}-ized algorithms
only using CUDA cores, but not Tensor Cores. \hl{Our implementations try to leverage the 
highly optimized functions from cuASR or CUTLASS whenever appropriate. Different from backend functions used in 
Section~\ref{sec:correctness_validate}, this version of code does not manually partition the algorithms based on 
our proposed \SIMDD{} hardware configuration but allow the code to fully exploit the performance from CUDA cores.}
This version helps us to identify the performance variance by naively applying
matrix algorithms without the presence of appropriate matrix accelerations. 

\paragraph{\SIMDD{} using Tensor Cores} This version of code use 
identical algorithms to the version of \SIMDD{} in CUDA cores except that we replace 
\hl{these algorithms' matrix operations to \SIMDD{} ones when appropriate. As existing hardware does not 
support our proposed \SIMDD{} operations yet, we evaluate the performance and validate the result of this 
version through the framework that Section~\ref{sec:emu_validate} describes.} 

\ignore{
\subsection{Experimental kernels}
As \SIMDD{} instructions are specifically targeting matrix-matrix problems that fulfill the 
computational pattern of $D = A \oplus (B \otimes C) $, along with previously mentioned operators(semirings), 
the software abstraction of \SIMDD{} APIs can be referred as figure~\ref{code:host}. In order to demonstrate 
the usage and behavior of such software abstraction, we designed/applied a total of 4 kernel functions with 
dedicated purpose to emulate the correctness, programmability, scalability and performance of \SIMDD{} instructions.
\input{codeHostKernel}
\subsubsection{Host kernel}
A single-threaded CPU function that directly implements the software abstraction of \SIMDD{} API. This kernel 
function offers the correctness of the computation result against other kernels.
\subsubsection{cuASR kernel} A high-performance GPU kernel provided by cuASR \cite{}, a CUDA library of Algebra for 
semirings. cuASR's backend leverages Nvidia cutlass\cite{?},  a high-performance matrix-multiplication library. 
cuASR kernels offers the state-of-the-art implementation of \SIMDD{} API using Nvida's GPU. This kernel can be
treated as software change to "implementation" of \SIMDD{} API.
\subsubsection{GPU simulated \SIMDD{} kernel} A simulation-purpose kernel implemented using CUDA, which 
demonstrates applying tiling algorithm with proposed 16x16x16 \SIMDD{} instructions. [could elaborate more here.]
\subsubsection{emulation kernel} A GPU-tensor core kernel leverages Nvidia cuBLAS API, which offers the potential 
performance of an MXU architecture implementing \SIMDD{} instructions. In contrast to the cuASR kernel, this can be 
treated as the hardware implementation of \SIMDD{} API.}
\ignore{
\subsection{Correctness and precision}
In order to demonstrate the correctness of \SIMDD{}, especially after applying the common optimization method of 
matrix multiplication. We applied each set of operators on the host kernel and the GPU simulated \SIMDD{} kernel 
to test the correctness on different shapes of matrices. We believe that the correctness benchmarks of the \SIMDD{} 
API are sufficient to demonstrate that 1. the \SIMDD{} API can inherit conventional matrix multiplication optimizations, 
whether for GPUs, CPUs, or special hardware. 2. the state-of-the-art GEMM kernel can be used to emulate the expected 
performance of the \SIMDD{} API. As a state-of-the-art software implementation of \SIMDD{}, the same correctness 
benchmark included cuASR kernels to ensure that the semiring matrix multiplication or semiring-liked matrix 
multiplication provided by cuASR is consistent with our \SIMDD{} API. Because the cuASR kernels were used 
extensively as designing performance benchmarking and implementing application of this work, we ensured the 
correctness of the cuASR kernels in the initial tests. Since the algorithms used in a large number of applications 
are in nature, only terminate at convergence, cuASR will be used as a correctness criterion for emulation and as a 
reference for the performance kernel to terminate its computation.

\subsection{Microbenchmark}
To test the performance of the \SIMDD{} API, we performed performance microbenchmarks for each set of operator's API. We compared the cuASR kernel with the emulation kernel for 8 different groups of operators to show the performance of \SIMDD{} with NVIDIA gpu-based architecture. Microbenchmarking can fully reflect whether \SIMDD{} has the conditions to exceed the traditional architecture with special hardware support.
To fully exploit the programmability and performance of our proposed programming model 
with \SIMDD{} APIs, we have selected a set of benchmark applications cross wide domains, 
solving problems included but not limited to the area of classical machine leaning and graph 
processing. Since our emulation method is based on Nvidia's GPU platform, for each benchmark 
application, we selected the state-of-the-art CUDA implementation along with best performing algorithms for SM based architectures. In this subsection, we will be introducing each baseline implementation and our emulation method with previously mentioned computation kernels and programming model.
\subsubsection{Operators and baseline applications}\hfill\\

}
Table~\ref{table:baseline}
lists the set of benchmark applications. Each of these applications
represents a use case for a proposed \SIMDD{} instruction as follows.
 
\noindent\textbf{All-Pairs Shortest Path (APSP) and All-Pairs Critical (Longest) Path (APLP)} 
APSP and APLP are graph problems that can be solved via
\texttt{min-plus} and \texttt{max-plus} \SIMDD{} instructions. Without \SIMDD{},
the most efficient implementation, \hl{ECL-APSP~\cite{ECL-APSP}}, applied a phase-based-tiled Floyd Warshall
algorithm to exploit massive parallelism using CUDA. 
We implemented APLP by extending the \andrew{\hl{ECL-APSP} with reversing the input weights on DAG} \ignore{APSP-CUDA~\cite{apspCUDA} codebase} to
support the desired recurrence relation. 
For \SIMDD{} version, the
implementation simply changes the function calls to use \texttt{min-plus} and \texttt{max-plus}. 
\ignore{
Among the many such implementations, we chose APSP-CUDA\cite{?} 
as the baseline implementation of min-plus. For max-plus, we inverse the edge value of the input 
graph, so that the baseline apsp algorithm can be directly applied to the APLP algorithm for directed 
acyclic graphs. It is worth mentioning that the max-plus semiring algebra, and in particular matrix 
multiplication using max-plus, is widely used in the matrix kleene star problem\cite{?}, which is 
nearly identical to the algorithm and data flow used in the APLP problem. There is reason to 
believe that max-plus has a wider range of applications.

\noindent\textbf{max-min, max-mul, min-mul:} Similar with min-plus and max-plus, applications aligned 
with max-min, max-mul, min-mul are also transitive-closure-liked graph processing problems. We extended 
the CUDA version of Floyd-Warshall algorithm \cite{?} by changing relaxation with corresponding \SIMDD{} 
supported operators. From the results, max-min, max-mul, and min-mul are used to implement the maximum 
capacity path, maximum reliability path, and minimum reliability path, respectively. The above algorithms 
are valuable in routing and networking related problems.
}

\noindent\textbf{Maximum Capacity Path (MaxCP), Maximum Reliability Path
(MaxRP) and Minimum Reliability Path (MinRP)} MaxCP, MaxRP and MinRP
represent another set of graph problems with solutions based on
transitive-closure. We select CUDA-FW as the state-of-the-art GPU baseline
for these problems and apply different operations in each iteration
of their algorithms. These applications' \SIMDD{} kernels simply
require invoking \texttt{max-min}, \texttt{max-mul} and \texttt{min-mul}
instructions. 

\noindent\textbf{Minimum Spanning Tree (MST)} Minimum spanning tree or minimum spanning forest
(MSF)
has rich applications in real-life network problems. However, conventional
MST or MSF algorithms cannot efficiently take advantage of GPU
architectures due to limited parallelism. The best-performing GPU
implementation that we know of is CUDA
MST and we use this one as our baseline. 
MST and MSF map perfectly to the \texttt{min-max} \SIMDD{} instruction. Our
\SIMDD{} version of code thus leverages \texttt{min-max} instruction to
investigate the efficiency of \SIMDD{} in this type of problem. 

\noindent\textbf{Graph Transitive Closure (GTC)} GTC is also a graph
analytics workload. Unlike
other graph algorithms, GTC simply checks the connectivity between all vertices
rather than reporting a route to fulfill the goal of optimization. Therefore,
GTC can use library functions from cuBool~\cite{cuBool} for efficient
implementation on GPUs. In \SIMDD{} version, we used \texttt{or-and}
instruction to implement the solution. 

\noindent\textbf{K-Nearest Neighbor (KNN)} Solving pair-wise L2 distance is
at the core of K-nearest neighbor and K-means problems, and can leverage
\SIMDD{}'s \texttt{add-norm} instruction. For the state-of-the-art GPU
baseline, we use KNN-CUDA. 

%% file: implementation.tex
\section{Emulation Setup}
\label{sec:emu}
To emulate the potential performance of \SIMDD{}, we reimplenmented the kernel function of a set of application mentioned in Section~\ref{sec:pm} using GPU-accelerated \SIMDD{} API (\textit{cuASR} kernel) and MXU-accelerated emulation API (\textit{cuBlas}). This section describes our experimental setup and the emulation method we developed.

\subsection{Experimental Setup}
Our hardware environment has 16 GB of main memory, which consists of a 8-core 16 threads AMD RyZen 3700X pocessor @4.4 GHz, which operates Ubuntu 16.04(Linux kernel version 4.15). A Nvidia RTX 3080 with 10 GB device memory built based on Nvidia's Ampere architecture is also installed in the host machine. Nvidia's third-generation Tensor Core come with GPU is used to emulate the performance of potential \SIMDD{} hardware. 
\subsection{Emulation Method}
In order to obtain the maximum performance of the tensor core, we chose to emulate the matrix operations supported by SIMD2 using \textit{gemmEX()} provided by cuBLAS. Since using simple matrix multiplication kernel is not enough to obtain correct results and we cannot control the convergence of the computation, we used the number of iterations run by the cuASR kernel to control the number of gemmex runs, which emulates the performance of selected application after using \SIMDD{} API.

%% file: tableBaseline.tex
\begin{table}[t]
\caption{Source and input data size of baseline implenmentation for each selected applications.}
\small

\vspace{-0.1in}
\begin{tabular}{|l|p{0.6in}|ll|}
\hline
\textbf{Application}                         & \multicolumn{1}{l|}{\textbf{Baseline Source}}                                                                      & \multicolumn{2}{l|}{\textbf{Input Dimension}} \\ \hline
\multirow{3}{1.2in}{All Pair Shortest Path\\(APSP)}    & \multirow{3}{0.7in}{ECL-APSP\\~\cite{ECL-APSP, apspforgpu}}                          & \multicolumn{1}{l|}{Small}       & 4096       \\ \cline{3-4} 
                                                    &                                                                                                                    & \multicolumn{1}{l|}{Medium}      & 8192       \\ \cline{3-4} 
                                                    &                                                                                                                    & \multicolumn{1}{l|}{Large}       & 16384      \\ \hline
\multirow{3}{1.2in}{All Pair Critical Path\\(APLP)}    & \multirow{3}{0.7in}{ECL-APSP\\ \cite{ECL-APSP, apspforgpu}}                          & \multicolumn{1}{l|}{Small}       & 4096       \\ \cline{3-4} 
                                                    &                                                                                                                    & \multicolumn{1}{l|}{Medium}      & 8192       \\ \cline{3-4} 
                                                    &                                                                                                                    & \multicolumn{1}{l|}{Large}       & 16384       \\ \hline
\multirow{3}{1.2in}{Maximum Capacity Path\\(MCP)}      & \multirow{3}{0.7in}{CUDA-FW\\ \cite{cudaFW, Lund2010AMC}}                             & \multicolumn{1}{l|}{Small}       & 4096       \\ \cline{3-4} 
                                                    &                                                                                                                    & \multicolumn{1}{l|}{Medium}      & 8192       \\ \cline{3-4} 
                                                    &                                                                                                                    & \multicolumn{1}{l|}{Large}       & 16384      \\ \hline
\multirow{3}{1.2in}{Maximum Reliability Path\\(MAXRP)} & \multirow{3}{0.7in}{CUDA-FW\\ \cite{cudaFW, Lund2010AMC}}                             & \multicolumn{1}{l|}{Small}       & 4096       \\ \cline{3-4} 
                                                    &                                                                                                                    & \multicolumn{1}{l|}{Medium}      & 8192       \\ \cline{3-4} 
                                                    &                                                                                                                    & \multicolumn{1}{l|}{Large}       & 16384      \\ \hline
\multirow{3}{1.2in}{Minimum Reliability Path\\(MINRP)} & \multirow{3}{0.7in}{CUDA-FW\\ \cite{cudaFW, Lund2010AMC}}                             & \multicolumn{1}{l|}{Small}       & 4096       \\ \cline{3-4} 
                                                    &                                                                                                                    & \multicolumn{1}{l|}{Medium}      & 8192       \\ \cline{3-4} 
                                                    &                                                                                                                    & \multicolumn{1}{l|}{Large}       & 16384      \\ \hline
\multirow{3}{1.2in}{Minimum Spanning Tree\\(MST)}      & \multirow{3}{0.7in}{CUDA MST~\cite{fastmemMST,CUDAMST,HARISH201277,PBBS,Thrust}} & \multicolumn{1}{l|}{Small}       & 1024       \\ \cline{3-4} 
                                                    &                                                                                                                    & \multicolumn{1}{l|}{Medium}      & 2048       \\ \cline{3-4} 
                                                    &                                                                                                                    & \multicolumn{1}{l|}{Large}       & 4096       \\ \hline
\multirow{3}{1.2in}{Graph Transitive Colsure\\(GTC)}   & \multirow{3}{0.7in}{CUBOOL\\ \cite{cuBool}}                                           & \multicolumn{1}{l|}{Small}       & 1024       \\ \cline{3-4} 
                                                    &                                                                                                                    & \multicolumn{1}{l|}{Medium}      & 4096       \\ \cline{3-4} 
                                                    &                                                                                                                    & \multicolumn{1}{l|}{Large}       & 8192      \\ \hline
\multirow{3}{1.2in}{K-Nearest Neighbor\\(KNN)}         & \multirow{3}{0.7in}{KNN-CUDA\\ \cite{knncuda}}                                        & \multicolumn{1}{l|}{Small}       & 4096       \\ \cline{3-4} 
                                                    &                                                                                                                    & \multicolumn{1}{l|}{Medium}      & 8192       \\ \cline{3-4} 
                                                    &                                                                                                                    & \multicolumn{1}{l|}{Large}       & 16384      \\ \hline
\end{tabular}
\label{table:baseline}
\vspace{-0.1in}
\end{table}

%% file: codeHostKernel.tex
\begin{figure}
\lstset{language=c}          
\centering
\begin{minipage}[b]{0.8\linewidth}
\lstset{language=C,label=SliceExaple}
\begin{lstlisting}[frame=single, numbers=left, mathescape,label=scopingExample]
for (i = 0; i < M; i++){
  for (j = 0; j < N; j++){
    for(k = 0; k < K; k++){
      D[i][j] = addop(C[i][j], 
        mulop(A[i][k], B[k][j]));
    }
  }
}
\end{lstlisting}
\end{minipage}
\caption{Software abstraction of \SIMDD{}}
\label{code:host}
\end{figure}

%% file: result.tex
\section{Results}
\label{sec:results}
This section summarizes our evaluation of \SIMDD{}. \SIMDD{} delivered up to
38.59\x{} speedup in benchmark applications with simply 5\% of total chip
area overhead. 

\input{area}

\begin{figure}[t]
    \begin{center}
    \includegraphics[width=1\columnwidth]{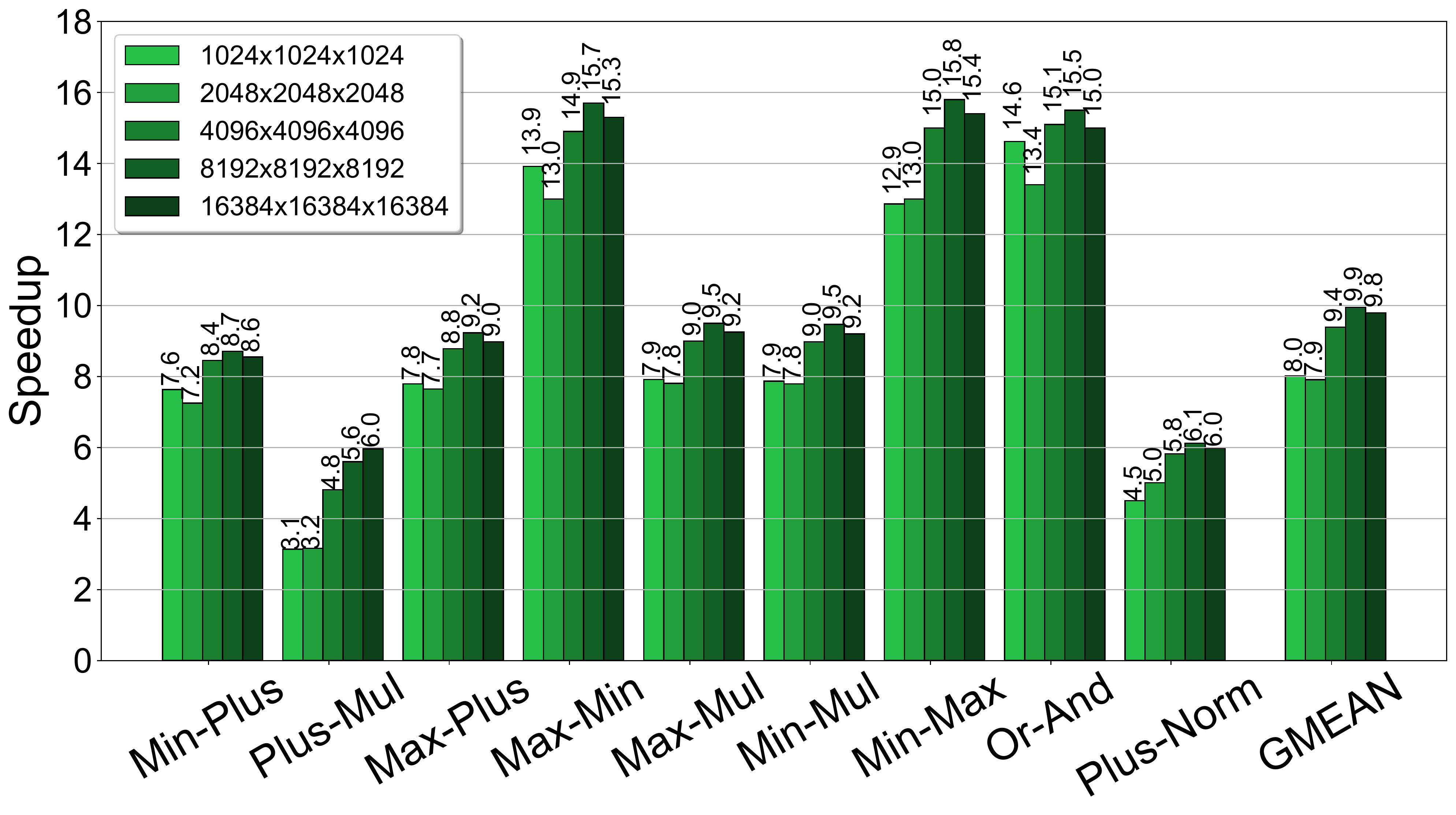}
    \vspace{-0.1in}
    \caption{Performance of microbenchmark with square matrices using \SIMDD{} API}
    \label{fig:micros}
    \end{center}
    \vspace{-0.1in}
\end{figure}

\subsection{Microbenchmarks}
\label{sec:microbenchmark}

\begin{figure}[t]
    \begin{center}
    \includegraphics[width=1\columnwidth]{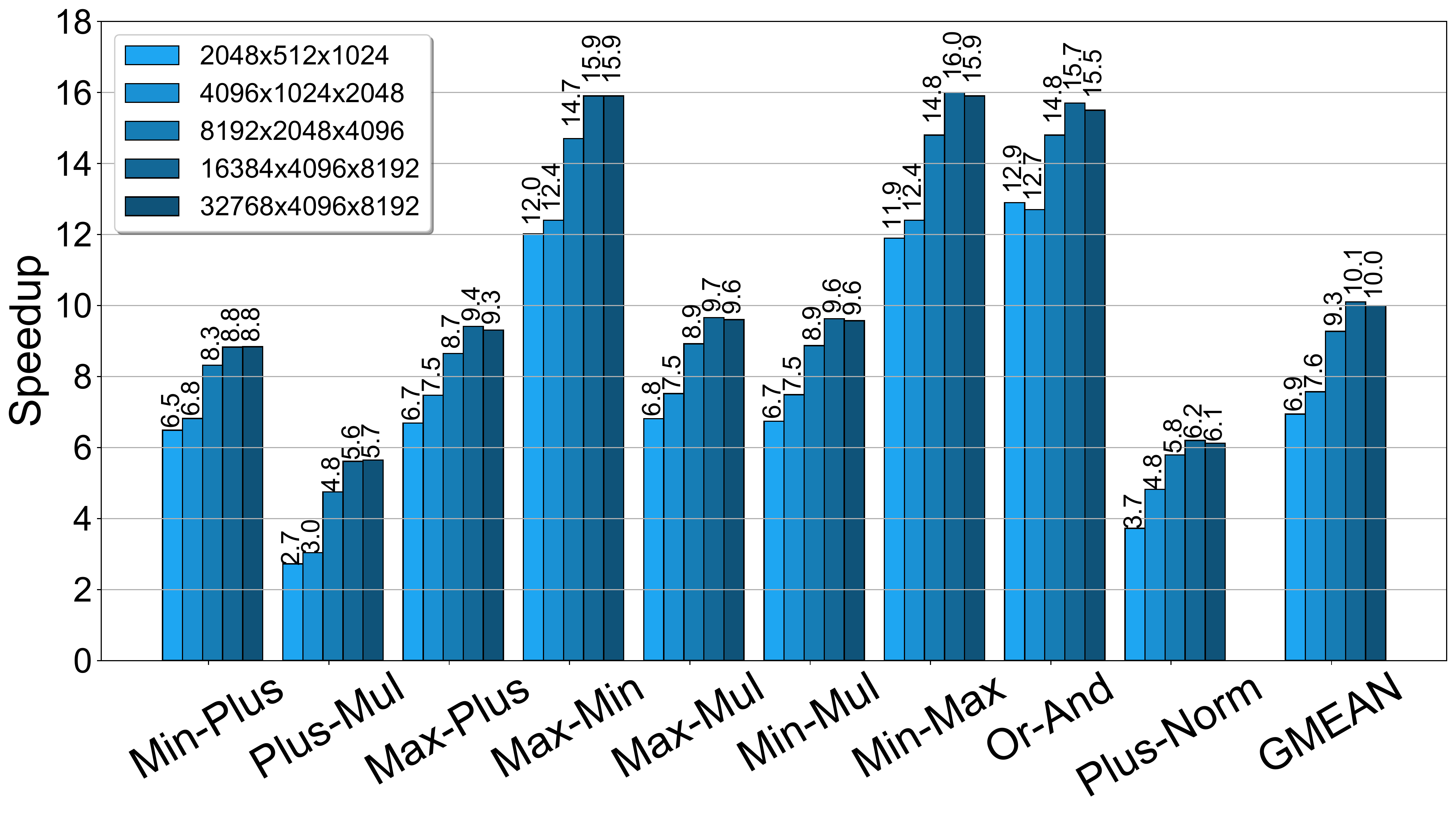}
    \vspace{-0.1in}
    \caption{\label{fig:microns}Performance of microbenchmark with nonsquare matrices using \SIMDD{} API}
    \end{center}
    \vspace{-0.1in}
\end{figure}

\begin{figure*}[t]
    \begin{center}
    \includegraphics[width=1\textwidth]{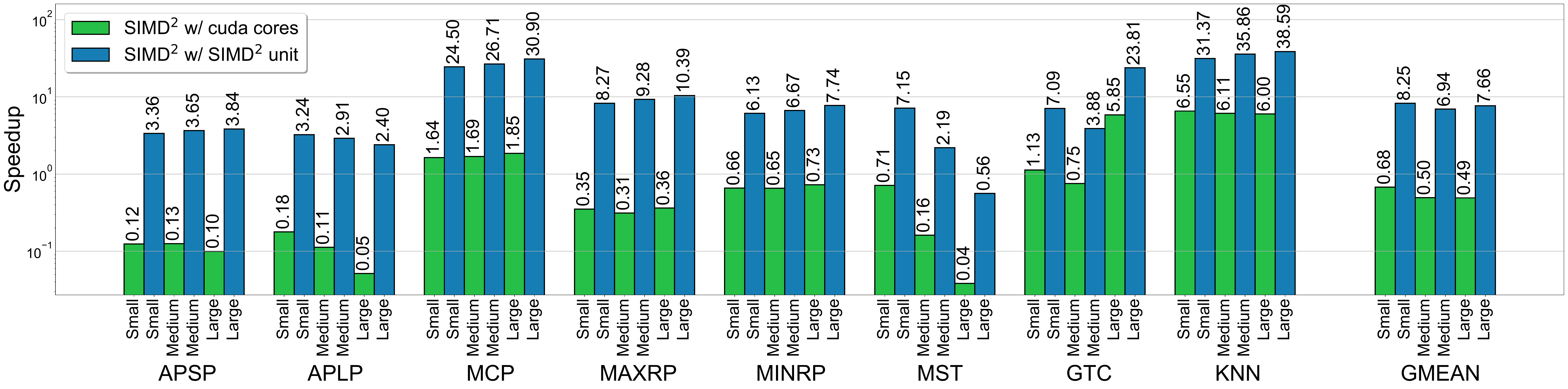}
    \vspace*{-0.2in}
    \caption{Performance of applications using \SIMDD{} API}
    \label{fig:app}
    \end{center}
\end{figure*}
We used microbenchmark workloads that repetitively invoke \SIMDD{}
the same instructions to gauge the performance gain of using \SIMDD{} units
compared against equivalent GPU implementations. The result shows up to
15.8\x{} speedup in evaluated scenarios. 

Figure~\ref{fig:micros} shows the performance gain of \SIMDD{} over the equivalent GPU
baseline implementations when using square matrices as inputs. \SIMDD{}
reveals up to 15.8\x{} speedup compared with using CUDA cores to achieve the
desired matrix operation on the same dataset. 
The geometric mean (gmean) that discounts the outlier
also shows a strong 7.9\x{}--9.9\x{} speedup, depending on the input set
sizes. When input matrices are larger than 4,096\x{}4,096 ones, the
performance gain saturates at about 10\x{}, representing the level of peak performance
gain of these instructions. 
Figure~\ref{fig:microns} shows the performance gain of \SIMDD{} instructions
on different shapes of matrices. The performance gain still saturates at the
level of 10\x{} when matrices are large, regardless of their shapes. 

From both results, \SIMDD{} has the largest 
performance gains for \texttt{min-max}, \texttt{max-min}, and \texttt{or-and} instructions, by
up to 15.8\x{}.
Such improvement is larger than the peak throughput difference between vector units and \SIMDD{} units.
We suspect the extra benefit from \SIMDD{} units is due to the structural hazard in the GPU SM architecture,
where \texttt{min} and \texttt{max} operations share the same hardware resources(e.g., ALU port),
and so are \texttt{or} and \texttt{and} operations.
By fusing these operations in a single instruction,
\SIMDD{} unit avoids this bottleneck and results in much higher speedup.
The speedups of \texttt{Plus-Mul} and \texttt{Plus-Norm}
operations are relatively low compared with others, but still enjoy
a 3.1\x{} speedup over using CUDA cores. This is because CUDA cores provide support 
for fused multiply-add (FMA) that allow the GPU to complete
\texttt{plus-mul} operations with a single instruction.
We expect that supporting more instructions similar to FMA would also provide
similar performance boost to the class of problems that \SIMDD{} addresses.
Nevertheless, \SIMDD{} still has a significant advantage, obtaining a speedup 
of up to 5.96\x{} for larger matrix operations.
We conclude that the \SIMDD{} architecture has larger potential than fusing more vector operations,
which we leave to future work.

\begin{figure*}[t]
    \begin{center}
    \includegraphics[width=\textwidth]{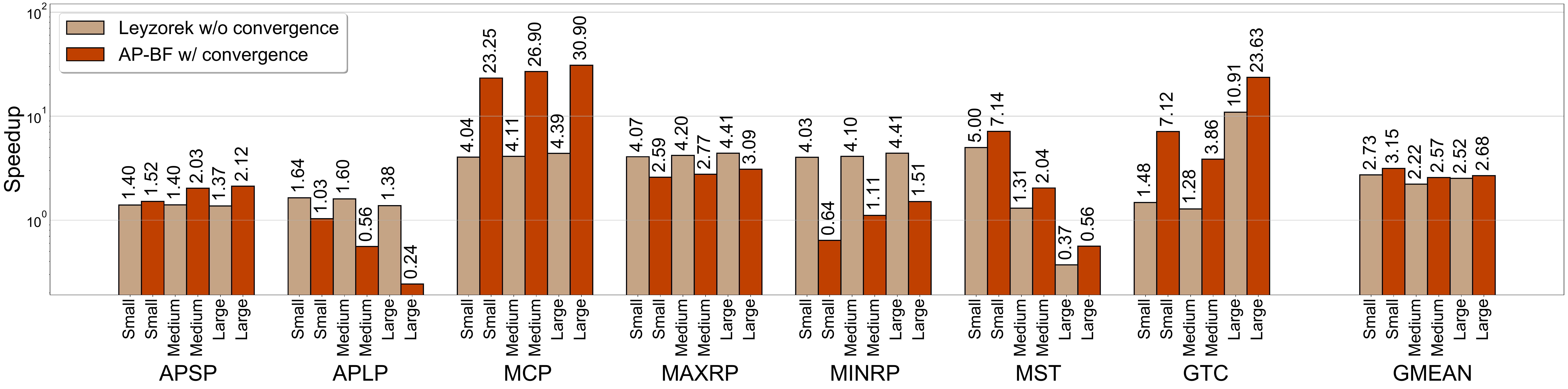}
    \vspace*{-0.2in}
    \caption{Performance of different algorithmic optimizations}
    \label{fig:conv}
    \end{center}
\end{figure*}

\begin{figure}[t]
    \begin{center}
    \includegraphics[width=\columnwidth]{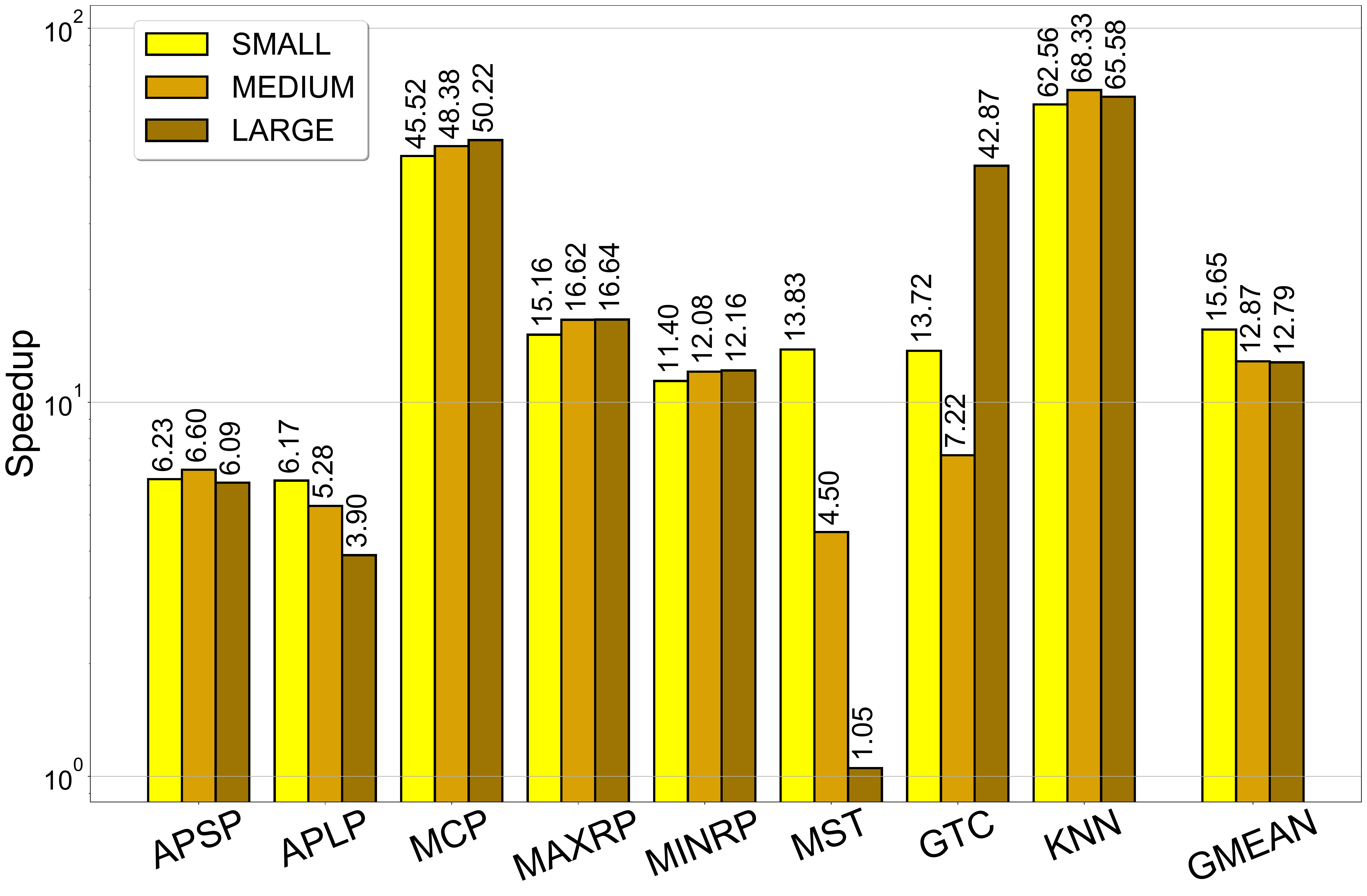}
    \vspace*{-0.2in}
    \caption{Performance of applications using Sparse \SIMDD{} unit}
    \label{fig:sparse_app}
    \end{center}
\end{figure}

\subsection{Benchmark Applications}
\label{sec:app_result}
Figure~\ref{fig:app} shows the speedup of kerenl latency of applications using \SIMDD{}
(\SIMDD{} w/ \SIMDD{} units) over the
baseline, optimized GPU implementations. \SIMDD{} achieves a geometric mean
of 6.94\x{} -- 8.25\x{}, with speedup as large as 38.59\x{}. 
The performance gain of \SIMDD{} in 7 out of the 8 applications remains strong
even when dataset sizes increased.

Compared with implementing the same matrix-based algorithms 
without \SIMDD{} presented (\SIMDD{} w/ CUDA cores), all applications show
significant slow down when \SIMDD{} units are absent. For APLP, MST, MaxRP,
MinRP, and APSP, these applications can never take advantage of
matrix-base algorithms due to their higher computational complexities
when \SIMDD{} units are absent. This result explains why these algorithms
were not favorable in conventional architectures. 
However, the introduction of \SIMDD{} 
makes these matrix algorithms feasible.
The matrix processing power from the \SIMDD{} unit can compensate or even 
improve the performance of the applications as our experimental results tell. 
In fact,
these algorithms can potentially take advantage of the embarrassingly
parallel nature of matrix multiplication to parallelize hard-to-parallelize
problems. 

\ignore{
Though the complexities of matrix algorithms used in \SIMDD{} are typically
not smaller than 
that of the classical algorithm as mentioned in the previous section, the 
performance improvement brought by the \SIMDD{} unit can compensate or even 
improve the performance of the applications from the experimental results. 
}
\ignore{
This result explains why
these algorithms were not favored in modern architectures. However, the
introduction of \SIMDD{} makes these matrix algorithms feasible. In fact,
these algorithms can potentially take advantage of the embarrassingly
parallel nature as matrix multiplication to parallelize hard-to-parallelize
problems. }

\ignore{
\hungwei{Don't understand what do you mean here. I still have questions
regarding GTC and MCP}
Our implenmentation using GPU accelerated \SIMDD{} achieved an average speedup of
1.36\x{}
}
\ignore{
\andrew{Group 1, cuASR $<$ baseline $<$ SIMD2}
Applications including APSP, MAXRP, and MINRP show significant slowdown when using 
only GPU acceleration due to the higher algorithm complexity, which shows why these 
algorithms were
not favored in modern architectures. However, because the number of iterations 
required for convergence is less, the performance gain from introduction of \SIMDD{} 
unit makes these matrix algorithms feasible.
}

For MCP, GTC, and KNN, their \SIMDD{} implementations out-perform their baseline,
state-of-the-art implementations, even without the presence of \SIMDD{}
units. For KNN, the computational complexity is the same for both \SIMDD{}
and the baseline implementations. However, the \SIMDD{} kernel can still achieve a maximum speedup of 
6.55\x{} without the
help of \SIMDD{} units. This is because the baseline implementation uses
customized functions to implement the algorithm, but the backend library of \SIMDD{}
without \SIMDD{} units leverages CUTLASS that is more optimized and
adaptive to modern GPU architectures. However, the performance gap between
configurations with or without \SIMDD{} units ranges between 4.79\x{} and
6.43\x{}. The performance advantage is more significant when we use the 
largest dataset. Therefore, even we revisit the design of the GPU baseline
and make that as efficient as \SIMDD{} on CUDA cores, such implementation
still has a huge performance gap to catch up with the performance using \SIMDD{} units.
For MCP and GTC, \SIMDD{} w/ CUDA cores can outperform their baseline
implementations even though the computational complexity is higher in
\SIMDD{} implementations for two reasons. The first reason is similar to the
case in KNN that \SIMDD{} w/ CUDA cores benefits from more optimized
library functions than the baseline ones. The other reason is that the rich
parallelism of these matrix-based algorithms allow these implementations to
scale better on modern GPU architectures -- considering that the RTX 3080 GPU
has twice as many CUDA cores than that of the previous generation of GPU architecture.
However, the state-of-the-art baseline implementation cannot take
advantage of this architectural improvement. On the other hand, this
result also reveals that \SIMDD{} programming model can make programs more
adaptive to various underlying architectures since these architectural
optimizations on \SIMDD{} operations will remain without the demand of
further code optimization.

The performance of APLP
and MST using \SIMDD{} degrades when datasets become larger.
This is because both APLP and MST using \SIMDD{} require 
additional convergence checks that are sensitive to input data values to determine the completion of the solution.
As the input dataset grows, the variance in the content also becomes more
significant and needs more iterations for the algorithm to converge.
However, if the number of iterations do not increase with the growth of dataset
sizes, the program can still show performance gain over conventional CUDA
cores since \SIMDD{} still makes each iteration faster. 
For MST, the baseline GPU solution uses Kruskal's algorithm that can solve 
MST/MSF problems with computational complexity at $O(E \log E)$~\cite{introToAlgo, Kruskalsalgo}, where $E$ is defined as the number of edges in 
the input graph. 
In contrast, each iteration of the matrix-based \SIMDD{} solution has the complexity
of $O(V^3)$~\cite{floydWarshall,introToAlgo}, where $V$ is the number of vertices in 
the input graph. Therefore, \SIMDD{} becomes slower than the baseline implementation
in each iteration for MST when dataset size is larger. 

\subsection{Discussion on algorithmic optimizations}
\ignore{
Our implementations in Section~\ref{sec:app_result} applied Leyzorek's algorithm and convergence checks
in all benchmark application except for KNN. 

\andrew{

Figure~\ref{fig:conv}
All-pairs Bellman-Ford algorithm with convergence check achieves and average of 6.15\x{} speedup, which is 
sub-optimal compared with results shown in Figure~\ref{fig:app} across all applications.
The worst case performance of using \SIMDD{} API achieved an average speedup of 3.35\x{}, which is emulated 
with applying Leyzorek's algorithm without convergence check.
}
}
In Figure~\ref{fig:app}, our implementations use Leyzorek's algorithm and convergence checks to optimize the number of SIMD$^2$ operations, except for KNN. As the proposed SIMD$^2$ architecture improves the performance of supported semiring-like operations, SIMD$^2$ still allows these matrix-based algorithms to outperform the baseline state-of-the-art GPU implementations without these algorithmic optimizations. 

The effect of convergence checks is sensitive to inputs. For each compute kernel using Leyzorek's algorithm on graph problems with $V$ vertices, the implementation will take $lg |V|$ SIMD$^2$ operations in the worst-case scenario. To evaluate the worst-case performance, we implemented a version of these applications without convergence checks. 
Figure~\ref{fig:conv} illustrates the performance of these implementations with bars labeled as Leyzorek w/o convergence. The baseline remains the same as Figure~\ref{fig:app}. Despite the increasing numbers of iterations, all applications still outperform their baseline GPU implementations, ranging from 1.11\x{} to 10.91\x{}. 

In Figure~\ref{fig:conv},  we also present implementations of these applications using the less efficient all-pair Bellman-Ford algorithm
(AP-BF w/ convergence). As Bellman-Ford algorithm can take up to |V| SIMD$^2$ operations, using Bellman-Ford algorithm can slow down APLP and MST when datasets become large. MINRP can never beat GPU implementations if we use Bellman-Ford algorithm-based implementations. However, the performance gain remains significant for other applications as the advantage of SIMD$^2$ architecture out-weighted the shortcomings of increased computational complexity.

\subsection{\SIMDD{} for Sparse Workloads}
\begin{figure}[t]
    \begin{center}
    \includegraphics[width=\columnwidth]{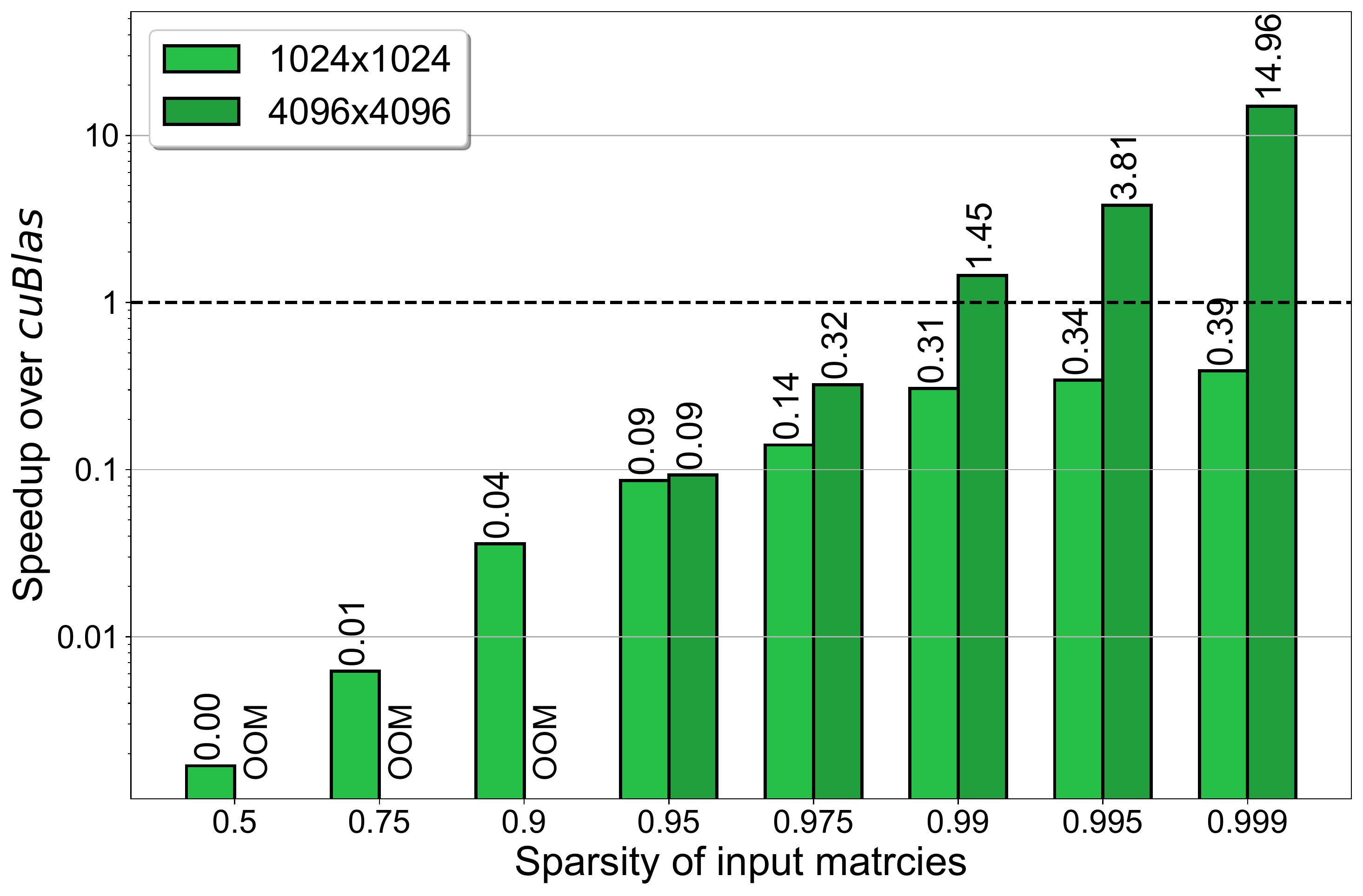}
    \vspace{-0.2in}
    \caption{\label{fig:spdense}Performance of sparse matrix multiplication}
    \end{center}
\end{figure}

\textbf{\SIMDD{} on architectural support for sparsity.} The idea of \SIMDD{} can be applied to architecture support for sparse inputs, too.
As an initial look of the \SIMDD{} model, we extend our emulation framework and build
on top of the \texttt{cuSparselt} library to model
the benefit of applying the \SIMDD{} idea to the sparse Tensor Cores in the RTX 3080 GPU,
which supports structured sparsity and provides up to 2$\times$ throughput.
We assume the inputs are pre-processed and stored in the format required by the sparse Tensor Core,
excluding the processing overhead when reporting the speedup.

Figure~\ref{fig:sparse_app} shows the speedup over baseline implementation when using a sparse \SIMDD{} unit.
We performed experiments using datasets with densities at 1\%. 
Using sparse \SIMDD{} units can improve performance by up to 68.33$\times$, 
with geometric means ranging from 12.79\x{}--15.65\x{}. Compared with 
the baseline \SIMDD{}, \SIMDD{} on sparse Tensor Cores is 1.67\x{}--1.9\x{} faster.

\hl{
\textbf{\SIMDD{} for extremely sparse inputs.}
Some applications often have extremely sparse inputs, especially for graph algorithms.
For these sparse inputs, a dense \SIMDD{} unit might provide less performance improvement
over implementations that are designed for sparse inputs, such as the \texttt{cuSparse} library.
We therefore explore at what sparsity \SIMDD{} can still provide benefits,
which is illustrated in Figure~\ref{fig:spdense}. The
x-axis in Figure~\ref{fig:spdense} shows the sparsity of inputs, meaning the
ratio of zeros to non-zeros in each dataset. The y-axis shows the speedup of using NVIDIA's
\texttt{spGemm} function, a sparse GEMM function optimized for Tensor Cores, from \texttt{cuSparse} library compared against
\texttt{gemmEx} function, a dense GEMM function for Tensor Cores, from \texttt{cuBlas} library. 
The results show that \textit{cuSparse} does not outperform \textit{cuBlas} for matrices of size $1024\times 1024$, 
and for matrices of size $4096\times 4096$, \textit{cuSparse} can outperform \textit{cuBlas} when the sparsity of the 
input matrix exceeds $99\%$.
This result shows that while many applications need to process sparse inputs,
there is still a range of sparsity where \SIMDD{} can provide benefit.
Such range also covers a number of real graph datasets that do not exceed the 
sparsity indicated in the results~\cite{GraphDensity},
implying that it is more efficient to use the dense matrix 
processing method for these cases if appropriate architectural support for sparse matrix
operations are absent.}

\hl{To handle extremely sparse inputs (sparsity $>=$ 99\%) on larger graphs,
we can apply \SIMDD{} sparse accelerators for spGEMM, which
also use multiply-and-add for the ALU, such as GAMMA~\cite{zhang:asplos:2021:gamma}.
For example, a GAMMA PE uses FP64 multiplier and adder, and an
\SIMDD{} GAMMA PE will use two FP64 ALUs, one supports the $\oplus$ op, and the
other supports the $\otimes$ op. This \SIMDD{} GAMMA accelerator would then be able
to run APSP on sparse graphs. 
In fact, extending sparse accelerators with \SIMDD{} would incur less 
overheads, as compute units contribute to less area than dense 
accelerators. For example, in GAMMA, only 10\% of the total area is due to the FP64 MAC unit.
We leave this extension to future work.
}

It is worth mentioning that while libraries like \texttt{cuSparse} have an advantage in terms of 
space complexity when dealing with extremely sparse matrices, the compressed matrix format may 
consume more device memory when storing relatively dense matrices. Experimental results show that 
\texttt{cuSparse} requires more memory than a single RTX 3080 GPU can provide when processing matrices with sparsity 
less than $90\%$ (the OOM result in Figure~\ref{fig:spdense}) and size more than $16384 \times 16384$.
However, 
when using a dense processing method, a 
GPU with 10GB of device memory can accommodate a matrix multiplication of at least $32768\times32768$ 
in size. 

%% file: area.tex
\input{tableArea}

\subsection{Area and Power}
\label{sec:area}
We implemented the proposed \SIMDD{} unit in RTL and synthesize them using
Synopsis design compiler and the 45nm FreePDK45 library. We extended a
baseline MMA unit that can simply perform MMA functions like
conventional MXUs presented in Tensor Cores. \hl{The baseline MMA unit features
4x4 matrix multiplications on 16-bit input elements and accumulates results 
in 32-bit elements. This configuration resembles the architecture used by 
Tensor Cores~\cite{nvidia-ampere} and Accel-Sim~\cite{Accel-Sim}.}
We carefully design the proposed extensions
to make the timing of the \SIMDD{} unit the same as the baseline. 
We empirically observe that our the modification for the \SIMDD{} unit never
increases the critical path delay.

Table~\ref{table:area}(a) lists the area overhead of adding \SIMDD{}
instructions
into the baseline MMA unit. The baseline MMA unit
is 11.52 $mm^2$ in size. Adding each individual instruction results in 1.34\% --
21.25\% overhead. The full-fledged \SIMDD{} unit has an area overhead
of 69.23\%. 
We inspected the public die photo of an NVIDIA 3080 GPU and 
found that SMs account for 50.2\% of the 628.4 mm$^2$ die area,
and each SM is 3.75 mm$^2$. If we scale the 69.23\% overhead from the 45nm 
process to the Samsung 8N process used for our 3080 NVIDIA GPU baseline,
a \SIMDD{} unit introduces only 0.378 mm$^2$, which is only 10\% of the SM 
area and 5\% of the total die area.

Table~\ref{table:area}(b) also lists the case where we only
implement a processing element to support a specific \SIMDD{} instruction
without the MMA function \hl{(i.e., as an individual accelerator). If we implement each \SIMDD{} instruction
separately as an individual accelerator, the total area of these accelerators
will require additional 2.96x space of the baseline MMA unit. In contrast, the design of \SIMDD{} unit allows these instructions
to reuse common hardware components and saves area. For example,}
we found that for the processing elements supporting Min-Mul and Max-Mul operations,
the area is almost the same as an MMA unit. However, combining their
functions into a single \SIMDD{} unit only results in 11.82\% of area
overhead, showing these instructions can share a large amount of circuits that
were originally used for MMA operations.
\hl{The baseline MMA unit consumes 3.74~W power. Extending the baseline as a
\SIMDD{} unit only adds 0.79~W to the active power. }

\hl{If we extend the baseline MMA to support 32-bit numbers, the size of the
MMA unit becomes 4.03x larger than a 16-bit MMA unit as
Table~\ref{table:area}(c) lists. A \SIMDD{} unit
supporting 32-bit inputs occupies 59\% more area than the 32-bit MMA unit.
If we further extend the MMA to support 64-bit numbers, the size of the MMA
unit becomes 11x larger than the 16-bit MMA. Extending the 64-bit MMA unit
as a 64-bit \SIMDD{} unit will add 52\% area overhead.}
\hl{If we make both the baseline MMA and \SIMDD{} units in supporting 8x8
matrix operations in 16-bit inputs, the MMA unit will become 7.5x larger
than the 4x4 baseline. The area overhead over the baseline MXU stays constant and scales well.}

%% file: tableArea.tex
\begin{table}[t]
        \small
        \centering
        \caption{The area overhead of supporting \SIMDD{} instructions through (a)
        adding instructions to the MMA unit, (b) individual accelerators, (c)
        extension to the MMA unit with various precisions, 
        compared to the baseline 16-bit MMA Unit. }
\begin{minipage}{.46\linewidth}
\centering
\begin{tabular}{|l|c|}
\hline
Supported Ops. & Area \\
\hline
MMA + All \SIMDD{} Insts. & 1.69 \\
\hline
MMA + Min-Plus & 1.21 \\
MMA + Max-Plus & 1.21 \\
MMA + Min-Mul & 1.12 \\
MMA + Max-Mul & 1.12 \\
MMA + Min-Max & 1.01 \\
MMA + Max-Min & 1.01 \\
MMA + Or-And & 1.04 \\
MMA + Add-Norm & 1.18 \\
\hline
\end{tabular}
\\(a)
\end{minipage}
\begin{minipage}{.46\linewidth}
\centering
\begin{tabular}{|l|c|}
\hline
Supported Ops. & Area \\
\hline
Min-Plus & 0.26 \\
Max-Plus & 0.26 \\
Min-Mul & 1.03 \\
Max-Mul & 1.03 \\
Min-Max & 0.06 \\
Max-Min & 0.06 \\
Or-And & 0.08 \\
Add-Norm & 0.19 \\
\hline
Total & 2.96 \\
\hline
\end{tabular}
\\(b)
\end{minipage}
\vspace{+0.1in}

\begin{tabular}{|l|c|c|c|c|}
\hline
         & 8-bit & 16-bit & 32-bit & 64-bit \\
\hline
MMA only & 0.25 & 1 & 4.04 & 11.17 \\
MMA + All \SIMDD{} Insts. & 0.69 & 1.69 & 6.42 & 17.01 \\
\hline
\end{tabular}
\\(c)
\label{table:area}
\end{table}

%% file: related_work.tex
\section{Related Work}
\label{sec:related_work}
In addition to the related work that motivates \SIMDD{} in Section~\ref{sec:caseofsimd2}, several
other lines of research that are relevant to \SIMDD{} deserve mention.
\subsection{Matrix extensions and instructions}
Instruction-level support for matrix-matrix multiplication can be dated back to the 90s.
MOM~\cite{corbal1999mom} proposes to leverage MXU to accelerate multi-media applications.
As neural networks become one of the most critical workloads,
commercial general processors now also include matrix instructions as well as MXUs to accelerate tiled-matrix-multiplication.
NVIDIA Tensor Core~\cite{T4,nvidia-ampere}, Intel AMX~\cite{IntelAMX}, and Arm SME~\cite{ArmSME}
all provide instructions for GEMM.
Our \SIMDD{} architecture is compatible with these prior work and modern designs.
\SIMDD{} reuses the existing hardware and software infrastructure to accelerate matrix operations beyond GEMM.

\subsection{Dense tensor accelerators}
\SIMDD{} builds on top of recent dense tensor accelerators for matrix-multiplication
~\cite{TPU,tpuv3,nvidia-ampere,qualcomm-ai100,huawei2021ascend,ibm-power10} to efficiently share data across datapath and reduce the bandwidth requirement of \SIMDD{} instructions.
While we implement our \SIMDD{} microarchitecture using systolic-array-like hardware structure,
other matrix-multiplication accelerator architecture,
such as the IBM MMA~\cite{ibm-power10} unit,
can be extended to support \SIMDD{} instructions.

In addition to matrix-multiplication, prior work also proposes accelerators for other dense linear algebra algorithms with different data sharing patterns.
For example, Weng et al.~\cite{weng2020hybrid} propose a hybrid systolic-dataflow architecture for inductive matrix algorithms (e.g., linear algebra solver).
Tithi et al.~\cite{tithi2014exploiting} propose a spatial accelerator for edit distance algorithms.
While these algorithms have a different data sharing pattern than \SIMDD{} instructions support,
we expect they can be implemented as \emph{CISC-\SIMDD{}} instructions with variable latency.
We nonetheless leave this extension to prior work.

\subsection{Sparse tensor accelerators}
\label{sec:sparse_tensor}
Since sparse matrices are common for many applications, such as HPC workloads, there is also ample prior work in sparse tensor accelerators 
~\cite{zhu2019sparse,hegde2019extensor,TensorPRAM, qin2020sigma, geng2020awb, abts2020think, srivastava2020tensaurus, srivastava2020matraptor, SparTen, zhang2020sparch, parashar2017scnn, zhou2018cambricon}.
These accelerators propose various sparse optimizations to skip ineffectual computations to speed up the tensor algorithms with sparse inputs.
They leverage various hardware support for gather/scatter operation and intersection to transform sparse tensor algebra into dense tensor algebra,
improving conventional dense tensor accelerators.
These techniques are therefore orthogonal to \SIMDD{}, and we expect \SIMDD{}
can be extended to support sparse tensor operation by applying similar
techniques, as discussed previously. 

\subsection{Graph algorithm accelerators}
While many graph algorithms can be expressed as tensor operations and linear algebra~\cite{graphblas} and accelerated by tensor accelerators,
prior work has also proposed hardware accelerators to speed up graph algorithms and analytics in their classic form.
Graphicionado~\cite{ham2016graphicionado}, GraphR~\cite{song2018graphr}, GraphP~\cite{zhang2018graphp}, and GraphQ~\cite{zhuo2019graphq} leverage processing-in-memory (PIM) architecture to alleviate the bandwidth bottleneck in graph algorithms.
PHI~\cite{mukkara:micro:2019:phi} and HATS~\cite{mukkara:micro:2018:hats} instead enhance conventional multi-core processors to accelerate common operations
in graph analytics, such as commutative reduction and traversal scheduling.
These hardware acceleration techniques focus on leveraging properties in graph algorithms to reduce data movement and bandwidth requirement.
In contrast, \SIMDD{} proposes a new instruction set for tensorized graph algorithms to leverage tensor accelerators ubiquitous in all compute platforms.

\subsection{Democratizing Domain-Specific Accelerators}
In addition to accelerating NNs, recent projects have demonstrated the strong potential of using NN/MMA accelerators for a broader spectrum of applications. Both Tensor Cores and TPUs can help improve the performance of linear algebra beyond GEMM~\cite{TCUSolver, GPTPU}, database queries~\cite{TCUSCAN, TPUDB, TCUDB}, cryptography~\cite{9715141} and scientific computing problems~\cite{EGEMM-TC, 9139823, lu2021large,9286192,9434068,9555937, 9563043,SimulationQuantumTPU}. Ray tracing accelerators are also useful for Monte Carlo simulations~\cite{9059266}, robotics navigation~\cite{9561068} and nearest neighbor search problems~\cite{RTNN}. However, due to the domain-specific nature of these accelerators, programmers have to intensively re-engineer the algorithm implementation to make use of these hardware accelerators. The resulting program may also incur overhead when transforming data structures to fulfill the demand of the target accelerator. By extending the hardware features, \SIMDD{} provides better programmability to reduce the overhead of remapping algorithms and allows applications that are not possible on conventional NN/MMA accelerators.

With hardware accelerators lifting the roofline, a critical issue is designing a memory hierarchy that streamlines the data inputs/outputs for computational logic. Potential solutions include bringing hardware accelerators closer to large memory arrays~\cite{TensorDIMM} or using other hardware accelerators to produce the demanding data structures for the target computing resource~\cite{NDS, TMA}.

%% file: conclude.tex
\section{Conclusion}
\label{sec:conclude}
Recent advance in hardware accelerators that accelerate matrix
multiplications in AI/ML workloads encourage us to take a new look at other
matrix problems. As many matrix problems share a similar computation pattern
with matrix multiplications that existing hardware accelerators already optimize
for, a more generalized matrix processor will allow these matrix problems to 
benefit from hardware acceleration. 

This paper introduces \SIMDD{} to investigate the potential of this research
avenue. We leverage the common computation pattern of significant matrix
problems to design the \SIMDD{} instruction set and implement 
a feasible, exemplary hardware architecture supporting these \SIMDD{}
instructions with 5\% total chip area overhead. We demonstrate the
effectiveness of \SIMDD{} using a set of benchmark applications, some of
them are rewritten with algorithms that are traditionally considered
inefficient due to the lack of hardware support like \SIMDD{}.
Our evaluation results show that the proposed \SIMDD{} architecture
achieves more than 6.94\x{} speedup on average across eight applications with various
tensor computation patterns.

%% file: acknowledgements.tex
\section*{Acknowledgments}
The authors would like to thank the anonymous reviewers
for their helpful comments.  
This work was sponsored by the two National Science Foundation (NSF) awards,
CNS-1940048 and CNS-2007124.
This work was also supported by new faculty start-up funds from University of California, Riverside. 

%% file: texfooter.tex


%

\bibliographystyle{plain}
\bibliography{paper}


\end{document}